\documentclass[10pt]{article}
\usepackage{authblk}
\usepackage{graphicx}
\usepackage{amsfonts}
\usepackage{float}
\usepackage{mathtools}
\usepackage[backend=bibtex]{biblatex}
\usepackage{subfigure}

\usepackage{amssymb,latexsym,amsmath,amscd,slashed,mathtools,mathrsfs,bm,stmaryrd,dsfont,amsthm}
\usepackage{setspace,scalerel,color,graphicx,enumerate}

\newcommand{\ostar}{\mathbin{\mathpalette\make@circled\star}}

\theoremstyle{plain}
\newtheorem{theorem}{Theorem}

\newtheorem{prop}[theorem]{Proposition}
\newtheorem{conjecture}[theorem]{Conjecture}
\theoremstyle{definition}

\DeclareMathOperator{\tr}{Tr}

\theoremstyle{plain}

\makeatletter

\date
\makeatother

\title{Bootstrapping the critical behavior of multi-matrix models}
\author[1]{Masoud Khalkhali}
\author[1]{Nathan Pagliaroli}
\author[2]{Andrei Parfeni}
\author[1]{Brayden Smith\footnote{\emph{Email addresses}:  masoud@uwo.ca, npagliar@uwo.ca, andrei.parfeni@yale.edu, bsmit288@uwo.ca}}
\affil[1]{Department of Mathematics, The University of Western Ontario}
\affil[2]{Department of Mathematics, Yale University}

\begin{document}
	\maketitle
	\begin{abstract}
		 Given a matrix model, by combining the Schwinger-Dyson equations with positivity constraints on its solutions, in the large $N$ limit one is able to obtain explicit and numerical bounds on its moments. This technique is known as bootstrapping with positivity. In this paper we use this technique to estimate the critical points and exponents of several multi-matrix models.  As a proof of concept, we first show it can be used to find the well-studied quartic single matrix model's critical phenomena. We then apply the method to several similar ``unsolved" 2-matrix models with various quartic interactions.  We conjecture and present strong evidence for the string susceptibility exponent for some of these models to be $\gamma = 1/2$, which heuristically indicates that the continuum limit will likely be the Continuum Random Tree. For the other 2-matrix models, we find estimates of new string susceptibility exponents that may indicate a new continuum limit. We then study an unsolved 3-matrix model that generalizes the 3-colour model with cubic interactions. Additionally, for all of these models, we are able to derive explicitly the first several terms of the free energy in the large $N$ limit as a power series expansion in the coupling constants at zero by exploiting the structure of the Schwinger-Dyson equations.
	\end{abstract}
	\tableofcontents
	\section{Introduction}

Matrix integrals are common objects of study in various areas of mathematics and physics. Originally, the first examples of convergent matrix integrals were introduced by Wigner \cite{wigner1958distribution}  to model the level-spacing of atomic nuclei. Further physical applications followed in quantum chaos and solid state physics \cite{guhr1998random,bleher2001random}.  Under certain general conditions,  matrix models have a $1/N$ expansion known as a genus expansion, where each term is a well-defined formal series in the coupling constants of the model that coincides with a generating function of combinatorial maps \cite{lando2004graphs}.   This was first discovered by `t Hooft in the context of gauge theory \cite{t1993two} and later developed further as models of string theory or quantum gravity  \cite{brezin1978planar,gross1991two,kazakov1985critical}. In particular, one is often interested in studying Hermitian matrix integrals that are of the general form
\begin{equation}\label{eq:general model}
		\mathcal{Z} = \int_{\mathcal{H}_{N}^{m}} e^{-\frac{N}{2}\sum_{i=1}^{m} \tr H_{i}^2 -N\tr S(H_{1},...,H_{m})}\prod_{i=1}^{m}dH_{i},
	\end{equation}
 where integration is with respect to the Lebesgue product measure on $m$ copies of the real vector space of $N\times N$ Hermitian matrices $\mathcal{H}_{N}$. The potential term $S(H_{1},...,H_{m})$ is a linear combination of words in the alphabet formed from the matrix variables $\{H_{1},H_{2},...,H_{m}\}$, that also contains some collection of coupling constants as coefficients. For simplicity, models where $S$ is an $m$-variate non-commutative polynomial are studied, although sometimes more general functions can be considered.

The primary quantities of interest in this paper are:
 \begin{itemize}
     \item the free energy $F=\ln \mathcal{Z}$,

  \item the (tracial) moments
	\begin{equation}\label{eq:tracicial moments}
		\langle\tr H_{i}^{\ell}\rangle = \frac{1}{\mathcal{Z}} \int_{\mathcal{H}_{N}^{m}} \tr H_{i}^{\ell}	 e^{-\frac{N}{2}\sum_{i=1}^{m} \tr H_{i}^2 -N\tr S(H_{1},...,H_{m})}\prod_{i=1}^{m}dH_{i},
  \end{equation}
	for any $i=1,2,...,m$ and $\ell \geq 0$, 
 \item and  the mixed (tracial) moments
		\begin{equation} \label{eq:mixed moments}
		\langle\tr W\rangle =  \frac{1}{\mathcal{Z}}\int_{\mathcal{H}_{N}^{m}}\tr W e^{-\frac{N}{2}\sum_{i=1}^{m} \tr H_{i}^2 -N\tr S(H_{1},...,H_{m})}\prod_{i=1}^{m}dH_{i},
	\end{equation}
	for any word $W$ in the alphabet of letters formed from the matrix variables $\{H_{1},H_{2},...,H_{m}\}$. 
  \end{itemize}
   The moments give us statistical information about the model and, in particular, if all moments are known, one can construct the distribution of eigenvalues \cite{jonsson1998spectral}. If the above model and quantities are interpreted as a formal series, they can be realized as the generating functions of connected maps with $m$-coloured edges \cite{lando2004graphs}.

     Typically, one says that a model is \textit{solved} (to leading order) if all these quantities can be written in terms of the coupling constants in the large $N$ limit. 
 In general, these models are notoriously difficult to solve, even to leading order. For single matrix models with polynomial potentials and a small class of multi-matrix potentials, subleading terms can be obtained recursively using a process known as Topological Recursion \cite{eynard2007invariants,eynard2016counting}.  For general multi-matrix models, very little is known and much of it lacks a rigorous basis, with the exceptions of works like \cite{guionnet2004first,guionnet2005combinatorial}. This is largely due to a shortage of analytic techniques. For an overview of what is known, see \cite{kazakov2000solvable,eynard2011formal}. 
 
 Of further interest is the study of the critical phenomena and a potential continuum limit of these models. Often, the quantities of interest have an asymptotic expansion around a critical point, i.e., a configuration of the coupling constants where one of the above quantities fails to be analytic. As a formal series that enumerates maps, such an expansion can be understood in terms of the asymptotic growth of the number of faces in the maps associated with the model. Such asymptotic expansions were first discovered in the 90's and are related   to 2D conformal field theory \cite{di19952d,witten1990two}. This connection comes from studying the statistical properties of planar maps as random metric spaces. Such spaces are viewed as piecewise flat geometries used to construct models of 2d Euclidean quantum gravity. One then hopes to take appropriate limits so that the size of each polygon tends to zero and that these random metric spaces converge in some sense to a continuum theory. In the past twenty years, much work has been done to rigorously study such continuum limits \cite{angel2003growth,chassaing2004random,le2010scaling}. Analogous ideas are used with Random Tensors \cite{bonzom2012random,lionni2018colored} and Random Dynamical Triangulations \cite{agishtein1992critical,loll2019quantum} for constructing toy models of higher dimensional gravity. It is worth noting that, despite matrix models corresponding to two-dimensional surfaces, multi-matrix models appear in both these frameworks as models with a space-time dimension greater than two \cite{ambjorn2001lorentzian,lionni2018colored}.

A strong indicator of the continuum behaviour of a model lies in its first non-integer exponent in the asymptotic expansion of the free energy in the large $N$ limit around a critical point 
 \begin{equation*}
     \lim_{N \rightarrow \infty} \frac{1}{N^2}\ln \mathcal{Z} \sim f(g-g_{c}) + (g-g_{c})^{2-\gamma}+... \quad \quad g \rightarrow g_{c},
 \end{equation*}
 where $f$ is a polynomial function and $2 -\gamma \not \in \mathbb{Z}.$ This expansion is often referred to as a double scaling limit. The constant $\gamma$ is 
known as the \textit{string susceptibility exponent} \cite{di19952d}.
When $\gamma = -1/2$, the associated continuum limit is expected to be the Brownian map \cite{le2013uniqueness,marckert2006limit}. When $\gamma = 1/2$, the associated continuum limit is expected to be the Continuum Random Tree \cite{aldous1991continuumI,aldous1991continuumII,aldous1993continuumIII} which is known as branched polymers in the physics literature \cite{cates1985fractal}. While these two are the most common and well-studied, other exponents are possible \cite{kazakov1989appearance,korchemsky1992matrix}. In particular, this is possible when the model studied has multiple coupling constants that can be fine-tuned \cite{distler19902d,ambjorn2016generalized}. Note that critical exponents do not always imply a continuum limit,  but rather give a heuristically strong indication of them. For an example, see \cite{kazakov1986ising}.

To combat the above lack of analytical methods to solve multi-matrix models, the method of \textit{bootstrapping with positivity} was first introduced in the innovative work \cite{Lin2020}. Through this method, one derives successively stricter bounds on the moments of matrix models by combining the relations of the Schwinger-Dyson equations with positivity constraints on the spectral measure. Solving for these bounds is a non-linear optimization problem. In particular, in \cite{kazakov2022analytic,zheng2023bootstrap} it was shown to be applicable to multi-matrix models that are unsolved, providing much-needed insight. In the same works, the relaxation bootstrap method was introduced, turning this non-linear problem into a linear one.  Even more recently in \cite{li2024analytic}, further progress has been made in studying the 2-matrix model from  \cite{kazakov2022analytic} by adding an ansatz for the structure of moments.  In \cite{perez2024loop}, positivity constraints on matrix integrals over the unitary group were used for bootstrapping.
 Bootstrapping with positivity has also been used in the previous work of the first two authors to study matrix models motivated by Noncommutative Geometry \cite{hessam2022bootstrapping}, ultimately leading to the recent analytic solution for many moments and the free energy of these models \cite{khalkhali2024coloured}. The idea of bootstrapping with positivity constraints has also been used in other areas of mathematical physics such as matrix quantum mechanics \cite{aikawa2022bootstrap,berenstein2022bootstrapping,berenstein2023semidefinite,bhattacharya2021numerical,tchoumakov2021bootstrapping}, lattice gauge theory \cite{kazakov2024bootstrap,kazakov2023bootstrap}, Feynman integrals \cite{zeng2023feynman}, the Ising model on the lattice \cite{cho2022bootstrapping}, and more.

  \subsection{Outline of main results}
  In this paper, we will apply the bootstrapping with positivity technique to study the critical points and exponents of several multi-matrix models. In \cite{Lin2020}, it was expressed as a hope of the author that bootstrapping with positivity would eventually be used to find new critical points and phenomena in multi-matrix models. In this paper, we do exactly that. As a proof of concept, we start off by using the bootstrap technique to estimate the critical point and exponent of the quartic single matrix model with known connections to 2d conformal field theory \cite{di19952d}.

  We then  study several closely related unsolved  2-matrix models of the form 
\begin{equation*}
	\mathcal{Z} = \int_{\mathcal{H}_{N}^{2}}\exp \left\{-N g \tr \left(\pm\frac{1}{4}(A^{4} + B^{4})  \pm \frac{1}{2} ABAB \pm  A^2 B^2\right) - \frac{N}{2}\tr (A^2 + B^2)\right\}dA dB
\end{equation*}
where $g$ is some real coupling constant. The main result is the conjectured form of the asymptotic expansion of these models near a critical point, from which we can deduce the associated string susceptibility coefficient of $\gamma =1/2$. This would potentially place these models in the universality class of the Continuum Random Tree.  For other configurations, we find estimates of string susceptibility exponents that are neither $\gamma =-1/2$ nor $\gamma =1/2$, potentially indicating a new continuum limit.

By exploiting the structure of the Schwinger-Dyson equations, we are able to give the first few terms of the free energy in the large $N$ limit as a power series  around zero of the coupling constant. For example, for the first two configurations we study, we find that 
\begin{equation*}
 \lim_{N \rightarrow \infty} \frac{1}{N^2}\ln \mathcal{Z}   = (F^{GUE}_{0})^2 - 2g + 9 g^2 - 72g^3 +  756g^4 - \frac{46656}{5} g^5 +  \mathcal{O}(g^{6})
\end{equation*}
where $F_{GUE}^{0}$ denotes the free energy of the Gaussian Unitary ensemble (GUE) in the large $N$ limit.
The process by which we produce this is iterative and elementary in nature. With sufficient computational resources, one could extend this expansion to an arbitrary number of corrections.

Lastly we study the following unsolved 3-matrix model
\begin{equation*}
	\int_{\mathcal{H}_{N}^{3}}\exp \left\{\frac{-N g}{3}\tr(A^3 + B^3 + C^3) - gN\tr (ABC + ACB) -\frac{N}{2}\tr (A^2 + B^2 + C^2)\right\}dA dB dC.
\end{equation*}
 In this model we begin to see evidence of critical behaviour similar to that of a cubic single matrix model. Just as in the 2-matrix models, we also compute the first few terms of a power series expansion for the free energy at $g=0$.

This article is organized as follows. In Section \ref{sec:prelim}, we give the necessary background on matrix models, their critical behaviour, the Schwinger-Dyson equations, and bootstrapping with positivity. Next, in Section \ref{sec: quartic}, we demonstrate how bootstrapping can be used to find the well-known critical point and exponent of the quartic single matrix model. In Section \ref{sec:2-matrix}, we study the 2-matrix models mentioned above, and lastly in Section \ref{sec:3matrix} we bootstrap the 3-matrix model.  In Section \ref{Sec: conclusion}, we summarize our work and its outlook. Examples of the Schwinger-Dyson equations and moments can be found for each model in the Appendices. These collections of equations are illustrative and not necessarily the entire list of equations used in bootstrapping. They contain clear patterns that may lead to hints at general analytical solutions.

\section{Preliminaries}\label{sec:prelim}
\subsection{Matrix integrals}
 In this paper, we restrict our interest to matrix integrals that are over some Cartesian power of the real vector space of $N\times N$ Hermitian matrices, denoted $\mathcal{H}_{N}$. There are two types of vastly different mathematical objects commonly referred to as a matrix integral or a matrix model. 

First, there are convergent integrals over spaces of matrices, usually defined by an exponentially decaying matrix function. One can use such matrix integrals to define probability distributions called matrix ensembles. Generally, (Hermitian) multi-matrix integrals are matrix integrals over some number of copies of $\mathcal{H}_{N}$, of the form
\begin{equation*}
    \mathcal{Z} =\int_{\mathcal{H}_{N}^{m}} e^{-N S(H_{1},...,H_{m})}\prod_{i=1}^{m}dH_{i},
\end{equation*}
where $S$ is some function such that the integral converges and each $d H_{i}$ is the Lebesgue measure on $\mathcal{H}_{N}$.  With $ \mathcal{Z}$ being a finite real number, we can define an associated matrix ensemble, sometimes called the Gibbs measure 
\begin{equation*}
    \frac{1}{\mathcal{Z}}e^{-N S(H_{1},...,H_{m})}\prod_{i=1}^{m}dH_{i}.
\end{equation*}

In addition to computing the partition function $\mathcal{Z}$, one is often interested in computing the moments of such measures (see equations \eqref{eq:tracicial moments} and \eqref{eq:mixed moments})
which can then be used to recover the distribution of eigenvalues \cite{pastur2011eigenvalue}. Most results concerning such ensembles are found after taking the matrix size to go to infinity, with very few results for finite $N$. Such integrals have been studied extensively in the one matrix case for mostly polynomial potentials \cite{johansson1998fluctuations,jonsson1998spectral,ercolani2003asymptotics}, with 
applications to orthogonal polynomials and Log-Gases \cite{deift2000orthogonal,forrester2010log}. Explicit results for multi-matrix integrals are far less common, partly due to the fact that most techniques concerning single matrix ensembles rely on invariance of the measure to make a dimensional reduction of the integral.

The second kind of matrix integral is called a \textit{formal matrix integral} which is  a formal series constructed by expanding all non-Gaussian terms of an expression like \eqref{eq:general model} around some coupling constants and then interchanging the order of integration and summation. Originally studied in \cite{brezin1978planar}, such formal series have connections to string theory, conformal field theory, quantum gravity, and combinatorics \cite{di19952d,Lando2004}. In particular, in \cite{brezin1978planar}, it was realized that such integrals were the generating functions of maps i.e. graphs embedded onto orientable surfaces, considered up to orientation-preserving graph homeomorphisms. Under general conditions, one can also consider such integrals as a formal Laurent series. We say a matrix model has a \textit{genus expansion} if for any word $W$ in the alphabet of matrix variables we can write 
\begin{equation*}
    \frac{1}{N}\langle \tr W \rangle = \sum_{g\geq 0}N^{-2g} m^{g}_{W}
\end{equation*}
and 
\begin{equation*}
    \frac{1}{N^2}\ln \mathcal{Z} = \sum_{g\geq 0} N^{-2g} F_{g},
\end{equation*}
where each $m^{g}_{W}$ and $F_{g}$ is a formal generating series of maps on surfaces of genus $g$  that does not depend on $N$. This is analogous to loop expansions in Quantum Field Theory. For more details we refer the reader to \cite{eynard2016counting}.

Despite being fundamentally different objects, both convergent and formal matrix integrals have infinite systems of recursive relations they satisfy called the Schwinger-Dyson equations (SDE). If a matrix ensemble has a well-defined formal counterpart, often one finds that both sets of moments satisfy the same SDE in the large $N$ limit, which under general conditions can be shown to have a unique solution \cite{guionnet2005combinatorial}. The matrix models studied in this paper appear to each have a unique solution to the SDE that satisfies positivity constraints on their moments. It is worth noting here that the positivity constraints come from the Hamburger moment problem whose solutions form a convex set. Hence, if there is such a solution, it is either unique or there are infinitely many \cite{schmudgen2020ten}. 

In this paper, we will exclusively study solutions of the SDE in the large $N$ limit. If one is only considering tracial moments from one matrix variable, we will denote them as
\begin{equation*}
    m_{\ell} := \lim_{N \rightarrow \infty} \frac{1}{N}\langle \tr H^{\ell}\rangle. 
\end{equation*}
Note that we can represent any word in two non-commuting matrices, as an integer sequence. For example, the word $AAAA BB A BBBB = A^{4} B^2 A B^4$, can be represented as $(4,2,1,4)$. We denote mixed moments coming from a word $W$,  by $m$ whose subscript is the associated sequence. For example the the moment for the word $AAAA BB A BBBB$ in th elarge $N$ limit is denoted $m_{4,2,1,4}$.

In this work, we will focus mainly on studying the critical behaviour of such models mentioned in the introduction. The critical exponent often gives a good indication of the continuum limit. In order to extract such an exponent, we need to first find a critical point where such an asymptotic expansion is possible. We define a \textit{critical point} of a matrix model as a configuration of the coupling constants where the free energy or a moment fails to be a real analytic function. It is often the case that the free energy and moments of many studied matrix models have algebraic or logarithmic singularities at their critical points. Such behaviour can be interpreted using random maps as the divergence of the moments of the number of vertices, edges, or faces. For more details we refer to the recent review \cite{budd2023lessons}.

	\subsection{The Schwinger-Dyson equations}
 Let $W$ be a $m$-variate non-commutative polynomial in $\{H_{1},...,H_{m}\}$. Via Stokes' formula, the following equality holds 
\begin{equation*}
\sum_{i,j=1}^{N}\int_{\mathcal{H}_{N}^{m}}\frac{\partial }{\partial H_{p}}\left(W \right)_{i,j}   e^{-\frac{N}{2}\sum_{i=1}^{m}\tr H_{i }^{2} -N\tr S(H_{1},...,H_{m})}\prod_{i=1}^{m}dH_{i} = 0.
\end{equation*}
 Expanding the left-hand side of this equation, one can derive recursive relations between moments. Such relations are referred to as the \textit{Schwinger-Dyson equations}. One can see that the choice of this particular polynomial in the integrand lends itself nicely to relations between moments. For example, if we set $m=1$, $W = H^{\ell}$, and the potential equal to 
\begin{equation*}
    -\frac{N}{2 } \tr H^2 + N  \sum_{j=3}^{d}\frac{t_j}{j}\tr H^{j}, 
\end{equation*}
the resulting SDE are 
\begin{align*}
    \langle N \tr H^{\ell+1}\rangle &= \sum_{j= 0}^{\ell-1} \langle \tr H^{\ell-1-j}\tr H^{j}\rangle
    + \langle N \tr H^{\ell}V'(H)  \rangle.
\end{align*}
Taking the large $N$ limit the covariance term factorizes and we arrive at
\begin{align*}
    m_{\ell+1} &= \sum_{j= 0}^{\ell-1} m_{\ell-1-j} m_{j}
    + \sum_{j =3}^{d}t_{j} m_{\ell+j} .
\end{align*}
One can explicitly compute all moments of this model and, if considered as a formal model, even $1/N$ corrections can be computed using a process known as Topological Recursion \cite{eynard2016counting}.

The SDE usually rely only on a finite set of moments and mixed moments to be solved, which can usually be deduced by utilizing a structural property of the SDE. In the above example, one needs precisely $m_{1}, m_{2},...,m_{d-2}$ to solve for all other moments. In general this set of initial conditions is much harder to find, which brings us to the following notion. For a given range of the coupling constants, the \textit{search space} of a matrix model is the minimum number of moments required as initial conditions for the model's SDE. In order for this concept to be well-defined, it is of course required that such a model has a solution. As we will see, there are ranges of coupling constants in the models we study such that the SDE have no solution. At the time of writing this article there are no results establishing the existence or size of the search space of multi-matrix models in any generality.  General conditions required for SDE to have solutions were studied in \cite{eynard2019solutions,guionnet2019asymptotics}.  It is also very possible for a system of SDE to have more than one solution. Conditions for the existence of a unique solution are discussed in \cite{guionnet2005combinatorial}.  For more details on the derivation of the Schwinger-Dyson equations we refer the reader to \cite{eynard2016counting,guionnet2019asymptotics}.  

\subsection{Bootstrapping with positivity }
The positivity constraints that can be derived for moments and mixed moments originate from the positivity of the spectral measure. This is easiest to see by starting with the Hamburger moment problem, which goes as follows: given a sequence of candidate real moments $(m_{0},m_{1},m_{2},...)$, does there exist a positive Borel measure $\mu$ on the real line whose moments correspond precisely to this sequence i.e.
	\begin{equation*}
		m_{n} = \int_{\mathbb{R}}x^{n}d\mu (x), \quad n =0,1,2,...?
	\end{equation*}
	One can prove that such a probability measure exists if and only if the Hankel matrix of moments is positive semi-definite:
	
	\begin{equation}\label{eq:Hankel 1 matrix}
	\left[\begin{array}{ccccc}
	1 & m_1 & m_2 & m_3 & \cdots \\
	m_1 & m_2 & m_3 & m_4 & \cdots \\
	m_2 & m_3 & m_4 & m_5 & \cdots \\
	m_3 & m_4 & m_5 & m_6 & \cdots \\
	\vdots & \vdots & \vdots & \vdots & \ddots
	\end{array}\right]\geq 0.
	\end{equation}
	In other words,
	$$
	\sum_{j, k \geq 0} m_{j+k} c_j \overline{c_k} \geq 0,
	$$
	for all sequences $\{c_i \}_{i}^{\infty}$ of complex numbers with finitely many non-zero elements.  For a proof and conditions on uniqueness, see [35].  Finite constraints can then be derived by taking the determinant of sub-matrices of the Hankel matrix, for example:
\begin{align*}
	0 &\leq m_{2}-m_{1}^{2}\\
 0 &\leq -m_{1}^2 m_{4}+2 m_{1} m_{2}
   m_{3}-m_{2}^3+m_{2} m_{4}-m_{3}^2\\
   0 &\leq m_3^4-3 m_2 m_4 m_3^2-2 m_1 m_5 m_3^2-m_6 m_3^2+2 m_1 m_4^2 m_3+2
   m_2^2 m_5 m_3\\+&2 m_4 m_5 m_3+2 m_1 m_2 m_6 m_3-m_4^3
   +m_2^2
   m_4^2+m_1^2 m_5^2-m_2 m_5^2\\
   &-2 m_1 m_2 m_4 m_5-m_2^3 m_6-m_1^2
   m_4 m_6+m_2 m_4 m_6.
\end{align*}

In the context of matrix models, the Hamburger moment problem amounts to finding a spectral measure such that its moments
	$$
	m_n = \frac{1}{N}\langle \operatorname{Tr} H^n \rangle , \quad n=1,2, \ldots, 
	$$
 satisfy the above positivity condition. Often, the spectral measure has a nice smooth and compactly supported density function $d \mu(x)=\rho(x) d x$ when we consider the moments in the large $N$ limit. For single matrix models, this function can often be found explicitly in the large $N$ limit \cite{deift2000orthogonal}. However, in general, finding all moments is a tall order, and instead one aims to compute enough moments to obtain an approximation of such a measure. The moments themselves also tell us useful information about the model as well as the enumeration of certain kinds of maps in the formal setting.

 \textit{Bootstrapping with positivity} refers to a process by which one combines the positivity constraints from the Hamburger moment problem with the relations given by the Schwinger-Dyson equations to derive explicit or numerical bounds on the moments of a matrix model. If the dimension of the search space of a model is small, one can derive simple expressions for the moments by solving the SDE up to some cutoff order, and then plugging them directly into the positivity constraints. The models studied in this paper all seem to have a search space dimension of one or two, making them ideal bootstrap candidates.

To derive positivity constraints for multi-matrix models, we must generalize these ideas. Consider a sequence of real numbers $\{m_{w}\}_{w \in \mathcal{A}}$ indexed by words $W$ in the alphabet  formed from  $\{H_{1},H_{2},...,H_{m}\}$. The sequence is called \textit{tracial}  if for any two cyclically equivalent words $w_{1}$ and $w_{2}$, we have that $m_{w_{1}} = m_{w_{2}}$. The necessary (but not sufficient) condition that a tracial sequence corresponds to mixed moments of a multi-matrix model is that the symmetric Hankel matrix $(m_{w_{1}^{*}w_{2}})_{w_{1},w_{2}}$ is positive semi-definite, i.e. 

 \begin{equation*}
			\sum_{i,j}  c_{i}\overline{c}_{j} \langle \tr W_{i}^{*} W_{j}\rangle \geq 0,
	\end{equation*}
 for all sequences of complex numbers $\{c_{i}\}^{\infty}_{i=1}$ with finitely many non-zero elements.
 
 Consider, for example, a 2-matrix model with matrix variables $A$ and $B$. All observables can be constructed in the real vector space spanned by the basis of lexicographically ordered words in $A$ and $B$. This gives us the tracial sequence $\{1, m_1, m_1,m_{2} , m_{1,1}, m_{2}, \ldots\}$. Then, the following Hankel matrix is positive semi-definite:
\begin{equation}\label{eq:Hankel 2 matrix}
\left[\begin{array}{ccccccc}
1 & m_{1} & m_{1}  & m_{2}  & m_{1,1} & m_{2} &\cdots\\
m_{1} & m_{2}  & m_{1,1}  & m_{3}  & m_{2,1}  & m_{1,2} & \cdots\\
m_{1} & m_{1,1} & m_{2} & m_{1,2}  & m_{1,1,1} & m_{3} & \cdots\\
m_{2}  & m_{3} & m_{2,1} & m_{4}  & m_{3,1}  & m_{2,2}  & \cdots \\
m_{1,1} & m_{1, 2} & m_{ 1,1,1} & m_{1,3} & m_{1, 2 1} & m_{1,1,2 } & \cdots\\
m_{2} & m_{2,1} & m_{3} & m_{2,2 }  & m_{2,1,1}  & m_{4} & \cdots\\
\vdots & \vdots & \vdots & \vdots & \vdots & \vdots &\ddots
\end{array}\right]\geq 0.
\end{equation}
For more details on the non-commutative Hamburger moment problem we refer the reader to \cite{burgdorf2012truncated}.

\section{The quartic model}\label{sec: quartic}
As an illustrative example, we will solve the quartic formal matrix model using bootstraps, since it has a known solution with which we may compare our estimates. Consider the following matrix model: 
	\begin{equation*}
		Z = \int_{\mathcal{H}_{N}}e^{-N\tr \left(\frac{1}{2}H^{2 } + \frac{g}{4}H^{4}\right)}dH.
	\end{equation*}

	As a formal model, the leading order contribution was first computed in \cite{brezin1978planar} and genus expansion corrections to any order can be computed explicitly via Topological Recursion \cite{eynard2016counting}.  As a convergent model, its leading order contribution can be found using orthogonal polynomials or with the equilibrium measure approach \cite{deift2000orthogonal}. The solution to the formal and convergent models coincides at least to leading order.
	
	The Schwinger-Dyson equations in the large $N$ limit are
	\begin{equation}\label{eq:SDEs_quartic}
		m_{\ell+1} = \sum_{k=0}^{\ell-1}m_{k} m_{\ell-k-1}+ g\, m_{\ell+3}.
	\end{equation}
	All the odd moments are zero and the explicit solution for even moments is
	\begin{equation*}
		m_{2\ell} = a^{\ell} \frac{(2\ell)!}{\ell! (\ell+2)!} (2\ell +2 - \ell a),
	\end{equation*}
where $a = -\frac{1}{6 g}(1 - \sqrt{1+12 g})$. In particular,
$$m_{2} = \frac{4a-a^2}{3}.$$

\begin{figure}[H]
			\centering
			\includegraphics[width=\textwidth]{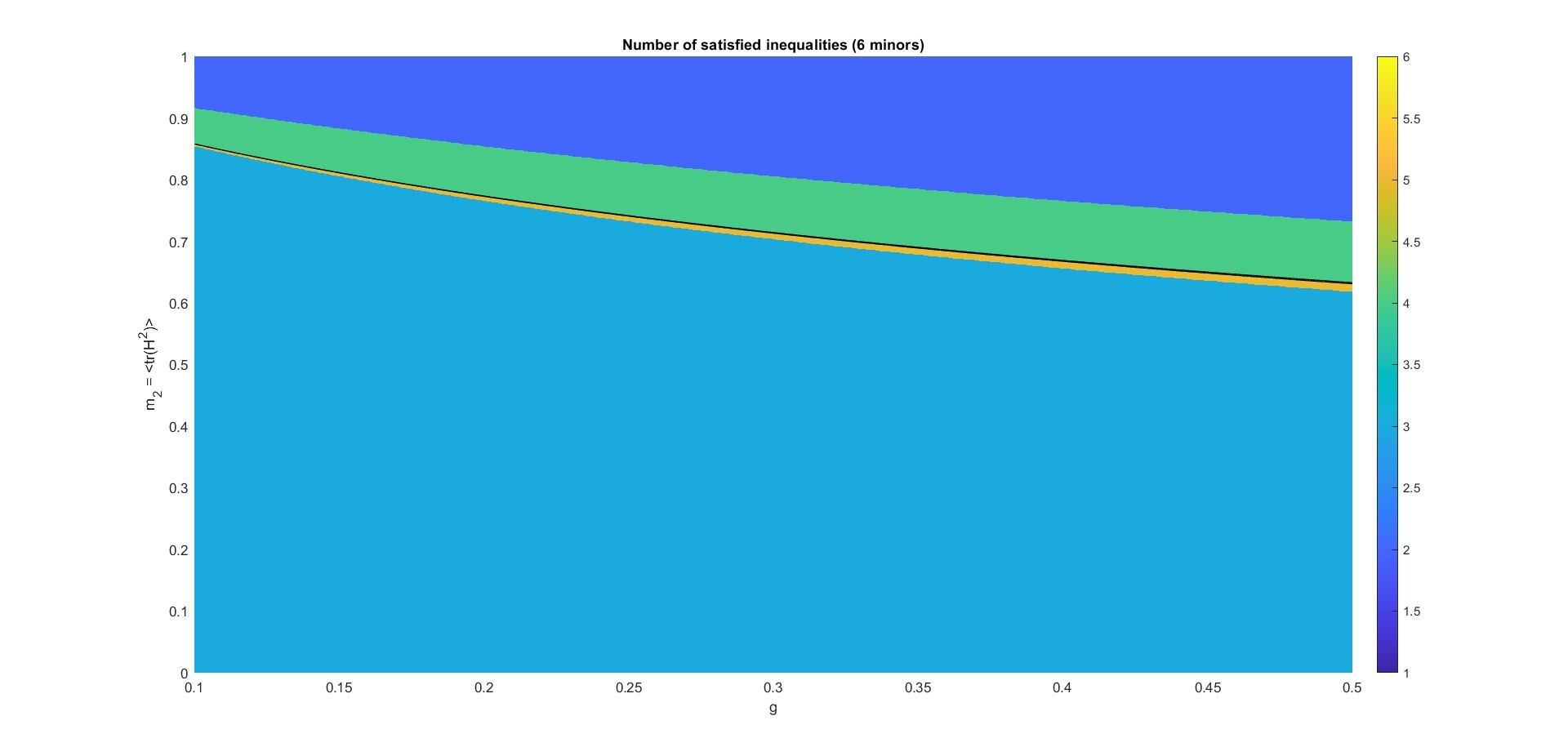}
			\caption{Bootstrapped solution of the quartic Hermitian matrix model for $g>0$. The colours correspond to the the size of the submatrix of the Hankel matrix as follows: Light blue is for 2 by 2, dark blue is for 3 by 3, green is for 4 by 4, and gold is for 5 by 5. The black curve is the analytic solution.}
			\label{fig:bootstrap quartic}
		\end{figure}

It is clear upon examination of the SDE \eqref{eq:SDEs_quartic} that the search space of these SDE has dimension one. In Figure \ref{fig:bootstrap quartic} one can see the bootstrapped solution for various sizes of Hankel matrices compared to the analytic solution. All plots in this section were generated in Matlab.

\begin{figure}[H]
     \centering
     \includegraphics[width=0.4\linewidth]{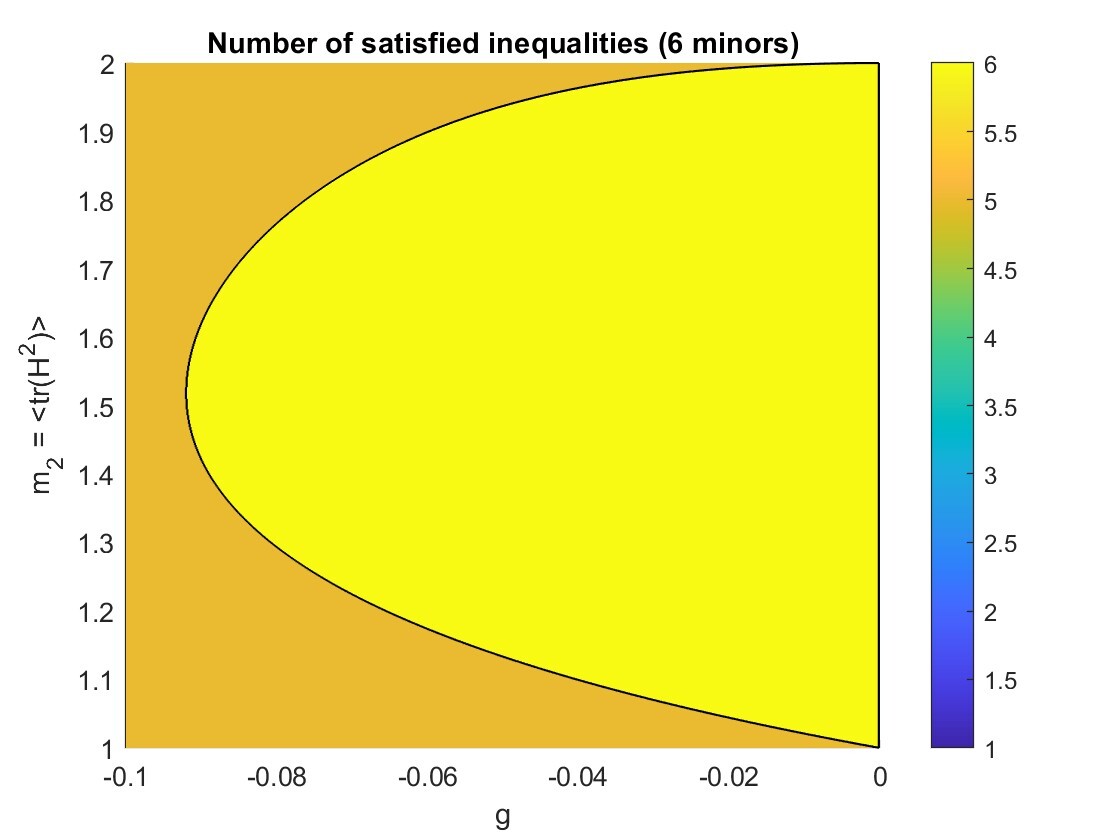}
     \includegraphics[width=0.4\linewidth]{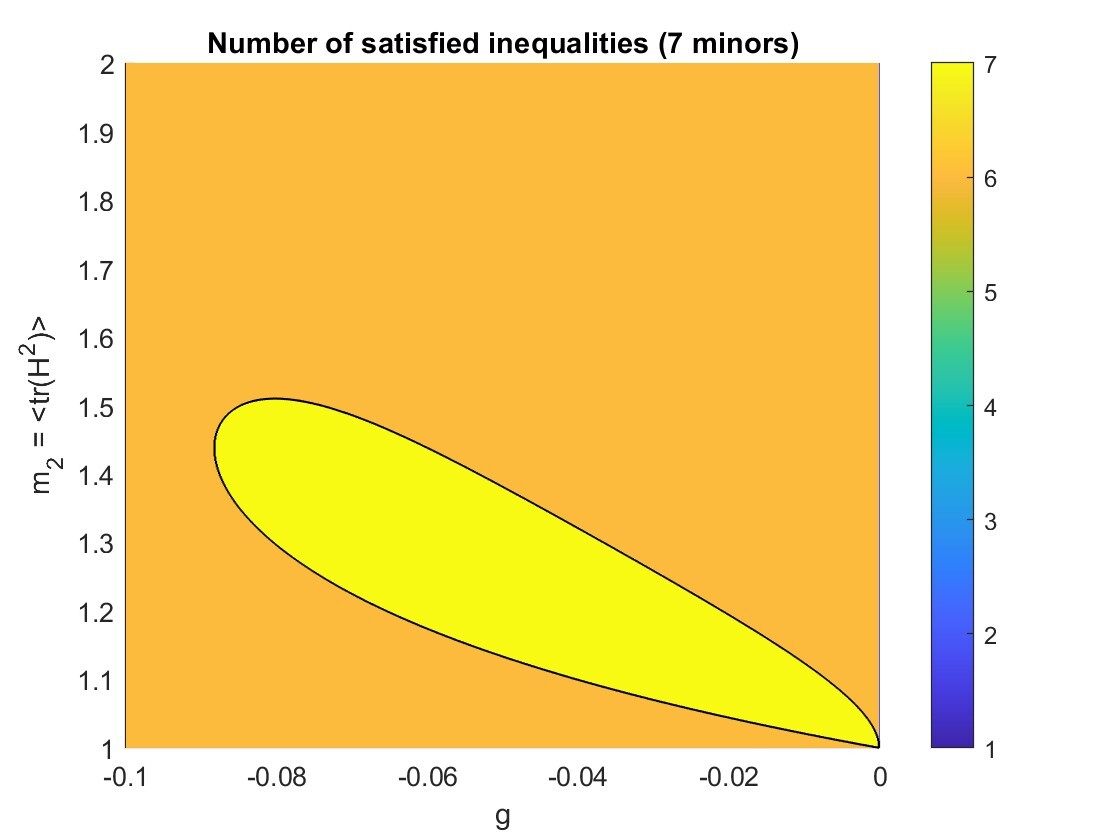}
     \includegraphics[width=0.4\linewidth]{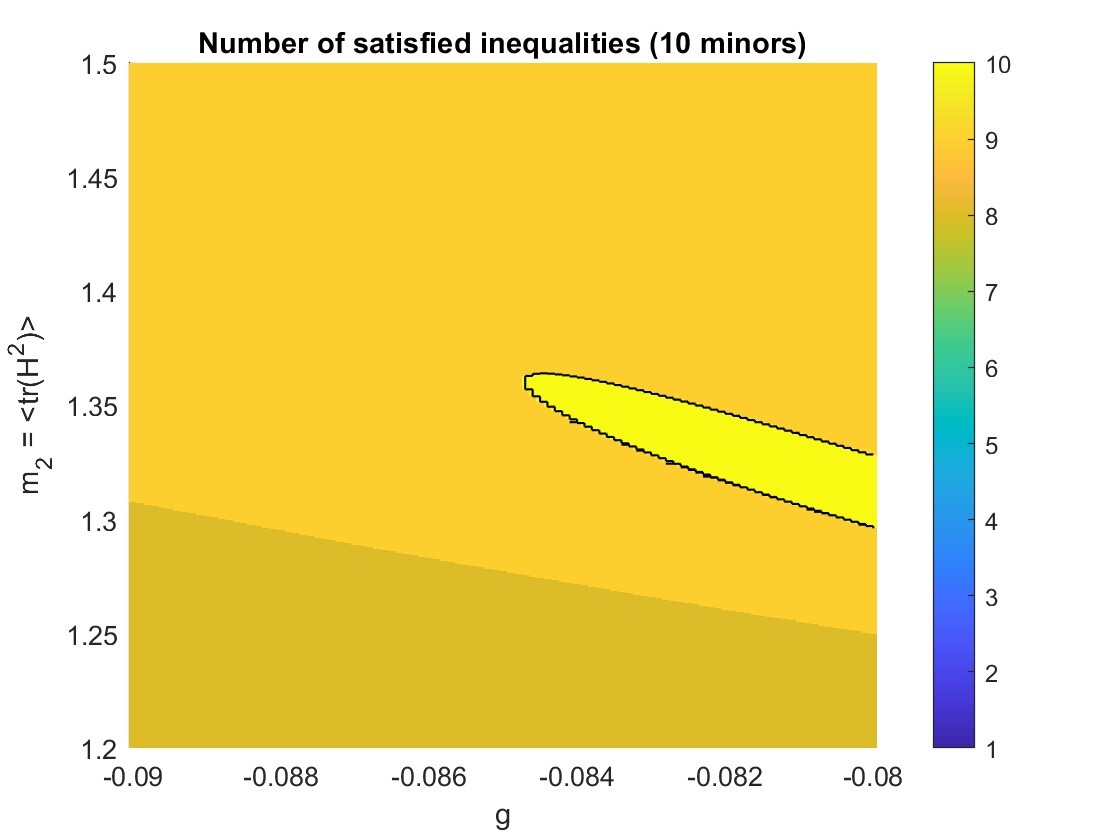}
     \caption{Bootstrapped solutions of the quartic Hermitian matrix model for $g <0$. The top left yellow region was computed with submatrices of size 6 by 6, the top right was computed with submatrices of size 7 by 7, and the bottom region was computed with submatrices of size 10 by 10. The critical point of this model is $-\frac{1}{12} = -.0833\overline{3}$.}
     \label{fig:bs_quartic critical}
 \end{figure}

The critical point of this model is $g_{c}= -\frac{1}{12}$. At this point one can recover the (3,2) minimal model from conformal field theory \cite{bergere2009universal}. Moreover, the bootstrapped solution converges to this point rather quickly, see Figure \ref{fig:bs_quartic critical}. This suggests that the bootstraps technique can potentially be used in general to find the critical points of matrix models with only a relatively small Hankel matrix size. Note that near the critical point we required larger submatrices to see convergence.

For illustrative purposes, we will derive the string susceptibility exponent of the model. We know that 
\begin{equation*}
    m_{2} \sim \frac{4}{3} - 16 (g-g_{c})  +64 \sqrt{3}  \left(g-g_{c} \right)^{3/2} \quad \quad  g \rightarrow g_{c}.
\end{equation*}
Taking the derivative, we have that
\begin{equation*}
    \frac{d}{d g} m_{2} \sim -16 + 96 \sqrt{3} (g-g_{c})^{1/2}\quad  \quad g \rightarrow g_{c}.
\end{equation*}
The second derivative of the free energy of the matrix model can then be expanded as 
\begin{equation*}
    \frac{d^2}{d g^2} F_{0} = \frac{d}{d g}\lim_{N \rightarrow \infty}\frac{1}{Z}\int_{\mathcal{H}_{N}} \frac{1}{4N} \tr H^{4} e^{-N\tr \left(\frac{1}{2}H^{2 } + \frac{g}{4}H^{4}\right)}dH=\frac{d}{d g} \frac{m_{4}}{4}\sim -4 + 24 \sqrt{3} (g-g_{c})^{1/2} \quad \quad g \rightarrow g_{c},
\end{equation*}
giving us the string susceptibility exponent of $\gamma = -1/2$.

\section{The 2-matrix models} 
\label{sec:2-matrix}

Consider the following 2-matrix model
\begin{equation}\label{eq:2-matrix model}
	\mathcal{Z} = \int_{\mathcal{H}_{N}^{2}}\exp \left\{-N \tr \left(\frac{g}{4}(A^{4} + B^{4})  + \frac{\alpha}{2} ABAB + \beta A^2 B^2\right) - \frac{N}{2}\tr (A^2 + B^2)\right\}dA dB.
\end{equation}
This model can clearly be considered as a convergent integral when the coupling constant $g,\alpha,\beta> 0$, or as a formal matrix integral that enumerates coloured maps otherwise. As far as the authors are aware, there is no  known technique that can be used to solve this model in its full generality, but it has been solved in special cases. When $(g,\alpha,\beta) = (g,0,0)$, this is two uncoupled quartic ensembles. When  $(g,\alpha,\beta) = (g,\alpha,0)$, this becomes the symmetric $ABAB$ model solved via the method of characteristic expansion in \cite{kazakov1999two}.  When $(g,\alpha,\beta) = (0,0,\beta)$, this matrix integral is sometimes known as the Hoppe model and has been solved via a reduction to the KP equation or saddle point method  \cite{kazakov1999d,berenstein2009multi}. For the particular configuration $(g,\alpha,\beta) = (g, \alpha,\alpha)$, this model was bootstrapped in \cite{kazakov2022analytic}, where a phase diagram was constructed.

In the following sections, we will examine special cases of this model for which the authors were able to study its critical phenomena. In particular, we are interested in estimating the location of critical points as well as the corresponding critical exponents of the moments and the free energy.  

All plots below were created in Mathematica, by first generating the Schwinger-Dyson equations in Python, then solving them using Mathematica's Solve[] function to identify the minimal generating set of moments (which provides important evidence about the search space) and the moment equations. Finally, we plot them using either Mathematica's RegionPlot[] or RegionPlot3D[] functions, depending on the conjectured dimension of the search space.

\subsection{When $(g,\alpha,\beta) = (g,g,g)$}\label{sec:ggg}

By symbolically solving the SDE of the model in Mathematica up to a cutoff, all the the mixed moments we examined could be expressed solely in terms of the coupling constant $g$ and $m_{2}$; see Appendix \ref{App:ggg_moments} for some examples. This suggests that this model has a search space of one. Despite the uncertain nature of this claim, by using the solutions mentioned above, we are able to produce several analytical results and conjectures in addition to numerical bootstrap estimates.

\subsubsection{Bootstrap bounds}
Note that the model is symmetric in $A$ and $B$, and that one could use the Hankel matrix \eqref{eq:Hankel 1 matrix} in either variable to derive constraints. In practice, we found that using constraints from the Hankel matrix \eqref{eq:Hankel 2 matrix} was less computationally expensive. In all regions of $g$ where this model was studied, the solution seems to converge to a square-root curve with a removable singularity at zero. In particular, using the five by five submatrix of the Hankel matrix \eqref{eq:Hankel 2 matrix} we are able to derive the following explicit bounds.

    \begin{prop}\label{prop:bound}
	The second moment $m_{2}$ of the matrix model \ref{eq:2-matrix model} is such that 
	\begin{equation*}
	-\frac{1}{4 g}	-\frac{1}{4} \sqrt{\frac{8 g+1}{g^2}}\leq m_{2}\leq -\frac{1}{4g}+ \frac{1}{4} \sqrt{\frac{8 g+1}{g^2}},
	\end{equation*}
	for $g \in [-1/8,0)$. Additionally,
	\begin{equation*}
		-\frac{1}{4} \sqrt{\frac{8 g+1}{g^2}}-\frac{1}{4 g}<m_{2}<0,
	\end{equation*}
and 
\begin{equation*}
0<m_{2}<\frac{1}{4} \sqrt{\frac{8 g+1}{g^2}}-\frac{1}{4 g}
\end{equation*}
for $g\in (0,\infty)$.
\end{prop}
 
\begin{figure}[H]
     \centering
     \includegraphics[width=0.4\linewidth]{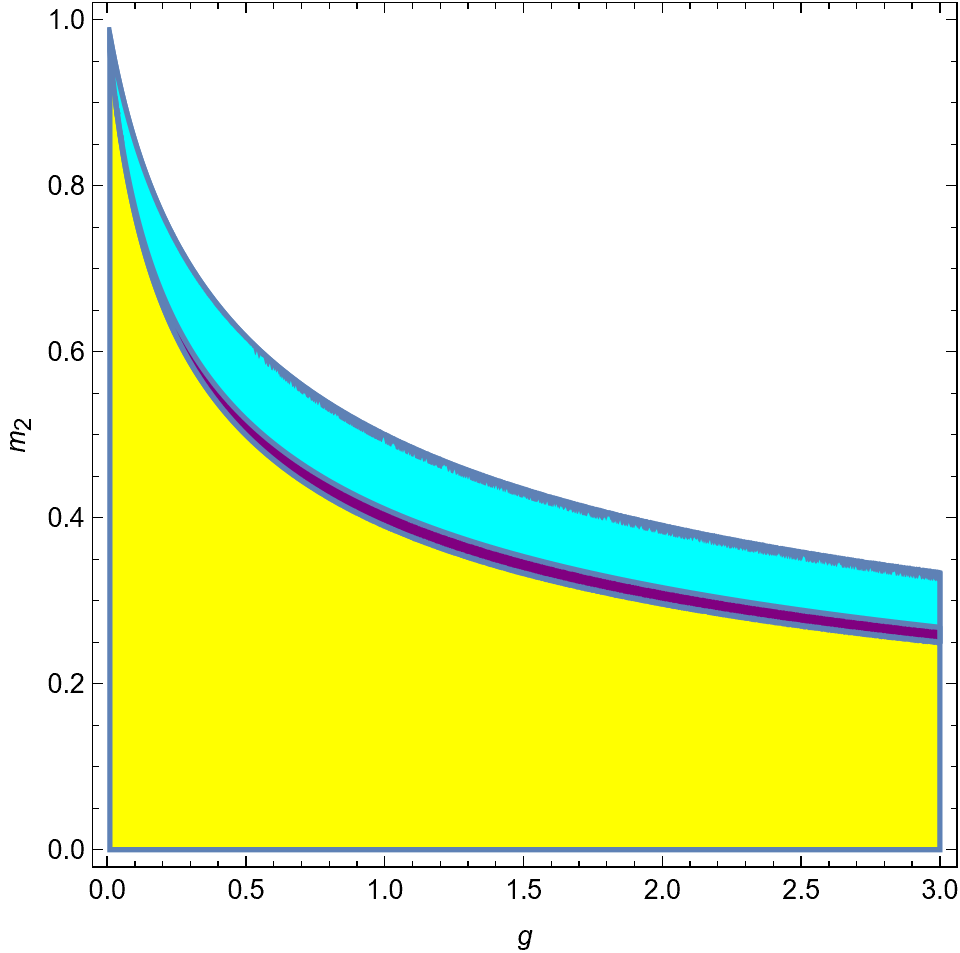}
     \includegraphics[width=0.4\linewidth]{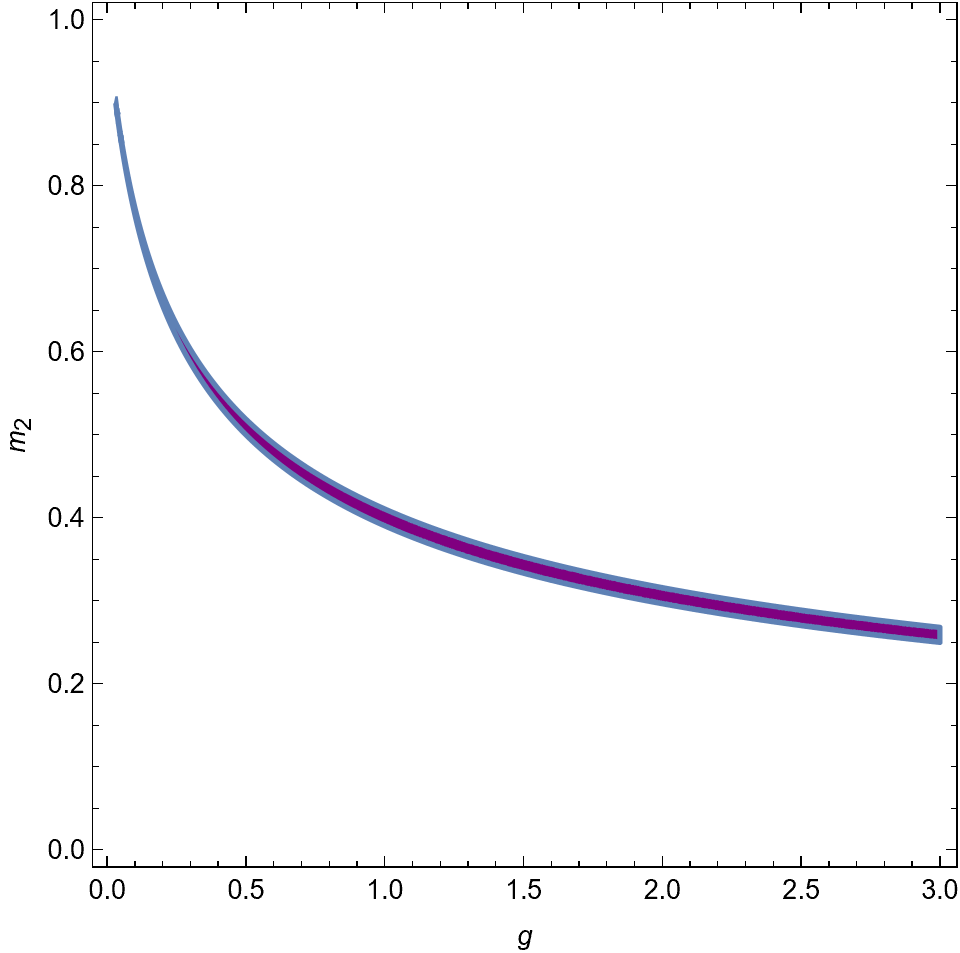}
     \caption{Bootstrapped estimates of $m_{2}$ of the $(g,\alpha,\beta) = (g,g,g)$ 2-matrix model. Each region corresponds to the positivity constraints of various sizes of submatrices of the Hankel matrix \eqref{eq:Hankel 2 matrix} overlaid as follows: yellow is 5 by 5, cyan is 9 by 9, and purple is 21 by 21. The subfigure on the right is only the latter constraint.}
     \label{fig:bs_112}
 \end{figure}

\begin{proof}
	These inequalities, among others, are derived by using the explicit expressions for  the moments 
	\begin{equation*}
		m_{4} = \frac{1 - m_{2} + 4 g m_{2}^2}{4 g},
	\end{equation*}
 and 
	\begin{equation*}
		m_{ABAB} =\frac{1 - m_{2} - 4 g m_{2}^2}{4 g},
	\end{equation*}
	with the determinant of the truncated Hankel matrix 
	\begin{align*}
		\det \begin{bmatrix}
			1   &  0 &  0 & m_2 & 0 & m_{2}  \\ 
			0 &  m_2 &  0 & 0 & 0 & 0\\
			0 &  0 &  m_2 & 0 & 0 & 0\\
			m_{2} &  0 &  0 & m_{4} & 0 & m_{2, 2}\\
			0 & 0 &  0 & 0 & m_{2, 2} & 0\\
            m_{2} & 0 & 0 & m_{2, 2} & 0 & m_{4}
		\end{bmatrix}\geq 0,
	\end{align*}
 giving us the inequality 
	\begin{equation*}
		g^2 m_{2}^2  ( m_{2}-1) ( 2 g m_{2}^2 +m_{2}-1) \geq 0,
	\end{equation*}
	from which the proposition follows.
\end{proof}
Similar explicit bounds can be obtained for the derivative of the free energy and for moments in terms of $m_{2}$ and $g$. This result and the proof serve as an example of how positivity constraints are to be derived for bootstrapping. For larger submatrices, the explicit bounds are far more complicated, but as we will see later, they are still approximated well by a square root curve. See Figure \ref{fig:bs_112} for the bootstrapped solution for $g \geq 0$. Note that the lack of purple region extending to zero is the result of numerical error of the solver caused by the potential removable singularity at zero.
 
    \subsubsection{Series expansion at zero}\label{sec:series expansion}
Another consequence of solving the SDE in terms of $m_{2}$ and $g$ is that we are able to deduce the first several terms in the series expansion of $\frac{d}{d g} m_{2}$ and $\frac{d}{d g}F_{0}$ as series at zero. This is done by using the formulae in Appendix \ref{App:ggg_moments} with the fact that the limit of all moments as $g$ goes to zero are the limiting moments of the GUE. For example, consider that 
\begin{equation*}
    \lim_{g\rightarrow 0} m_{4} = \lim_{g\rightarrow 0}\frac{1 - m_{2} + 4 g m_{2}^2}{4 g} = 2.
\end{equation*}
Applying L'H\^{o}pital's rule and rearranging, we find that
\begin{equation*}
    \lim_{g\rightarrow 0} \frac{d}{d g}m_{2} = -4.
\end{equation*}
The exact same procedure can be used to recursively find the derivatives of $m_{2}$. Up to the first four orders, we have computed the following:

    \begin{equation*}\label{eq: m2_der_expansion}
   \frac{d}{dg} m_{2} = -4 + 72 g -1296g^2 + 24192 g^3 - 466560 g^4 + \mathcal{O}(g^{5}),
\end{equation*}
or
 \begin{equation}\label{eq: m2_expansion}
   m_{2} = 1 - 4g + 36 g^2 - 432g^3 + 6048 g^4 - 93312 g^5 + \mathcal{O}(g^{6}).
\end{equation}

Now, we may write the free energy in terms of our moments as follows. Note that by differentiating under the integral sign, the derivative of the free energy of the model with respect to the coupling constant can be expressed as 
	\begin{equation*}
		-\frac{d}{d g}\ln \mathcal{Z} = \frac{1}{4} \langle \tr A^{4} + B^{4}\rangle + \frac{1}{2} \langle \tr ABAB \rangle  + \langle \tr A^2 B^2\rangle . 
	\end{equation*}
In the large $N$ limit, this will become 
\begin{equation*}
		-\frac{d}{d g}F_{0}= \frac{1}{2} m_{4} + \frac{1}{2} m_{ 1,1,1,1}  +  m_{2,2}. 
	\end{equation*}
Applying the equations for these moments from Appendix \ref{App:ggg_moments}, we have that the derivative of the free energy reduces to
\begin{equation*}
    \frac{d}{d g}F_{0} = \frac{m_{2}-1}{2g},
\end{equation*} from which we can deduce that
\begin{equation}\label{eq:free energy 2-matrix}
\lim_{N \rightarrow \infty}\frac{1}{N^2}\ln \mathcal{Z}   = (F^{GUE}_{0})^2 -2g +9 g^2 -72 g^3 + 756 g^4 - \frac{46656}{5}  g^5  +\mathcal{O}(g^{6}).
\end{equation}

If this solution is to a formal matrix integral of the form of \eqref{eq:2-matrix model}, then each term of this expansion has a graphical interpretation as the Boltzmann weight of maps glued from quadrangles whose edges are of one of two colours \cite{lando2004graphs}.

    \subsubsection{Critical behaviour}
\begin{figure}[H]
    \centering
    \includegraphics[width=0.5\linewidth]{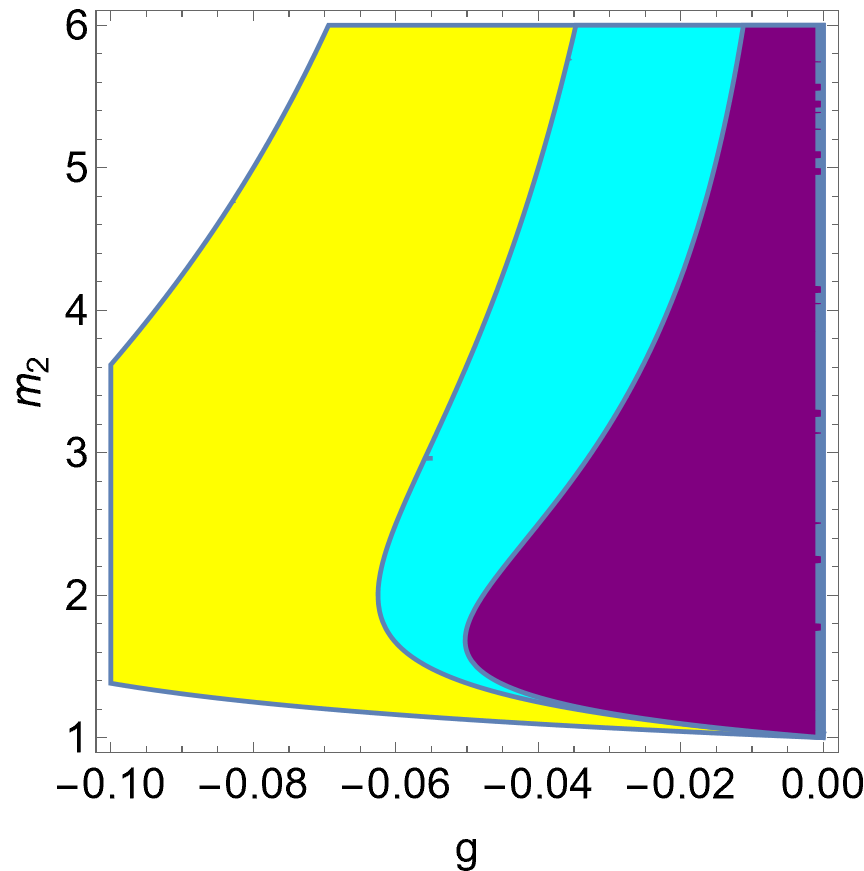}
    \caption{Bootstrapped estimates of $m_{2}$ of the $(g,\alpha,\beta) = (g,g,g)$ 2-matrix model for $g<0$. Each region corresponds to the positivity constraints of various sizes of submatrices of the Hankel matrix \eqref{eq:Hankel 2 matrix} as follows: yellow is 5 by 5, cyan is 9 by 9, and purple is 21 by 21.}
    \label{fig:bs_critical_112_1}
\end{figure}

    For small values of the submatrix size, we begin to see convergence to a potential critical point around $0>g>-0.05$, see Figure \ref{fig:bs_critical_112_1}. Using the eight by eight submatrix of the Hankel matrix \eqref{eq:Hankel 1 matrix}, the critical point estimate was found to be -0.0892435.  Estimates of -0.065 and -0.0502729 were found for nine by nine and twenty-one by twenty-one submatrices of the Hankel matrix \eqref{eq:Hankel 2 matrix}, respectively. 
    
 \begin{figure}[H]
    \centering
  
    \includegraphics[width=0.5\linewidth]{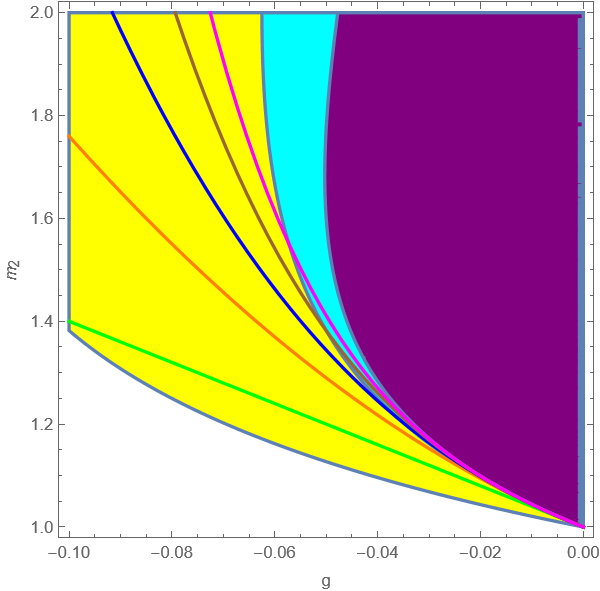}
    \caption{Bootstrapped estimates of $m_{2}$ of the $(g,\alpha,\beta) = (g,g,g)$ 2-matrix model for $g<0$. Each region corresponds to the positivity constraints of various sizes of submatrices of the Hankel matrix \eqref{eq:Hankel 2 matrix} as follows: yellow is 5 by 5, cyan is 9 by 9, and purple is 21 by 21. This is compared with the estimates from the series expansion of the second moment in Section \ref{sec:series expansion}. Each coloured line from left to right represents successively more terms in expansion \eqref{eq: m2_expansion}. }
    \label{fig:bs_critical_112_2}
\end{figure}

Moreover, our power series expansion of $m_2$ as a function of $g$ at zero in equation \eqref{eq: m2_expansion} allows us to plot the successive approximations of this function by polynomials in $g$. Note that if the alternating zero trend that appears in the expansion \eqref{eq:free energy 2-matrix} continues, then for $g<0$, each term in the series is positive. This can be seen in Figure \ref{fig:bs_critical_112_2}.

\begin{figure}[H]
    \centering
    \includegraphics[width=0.5\linewidth]{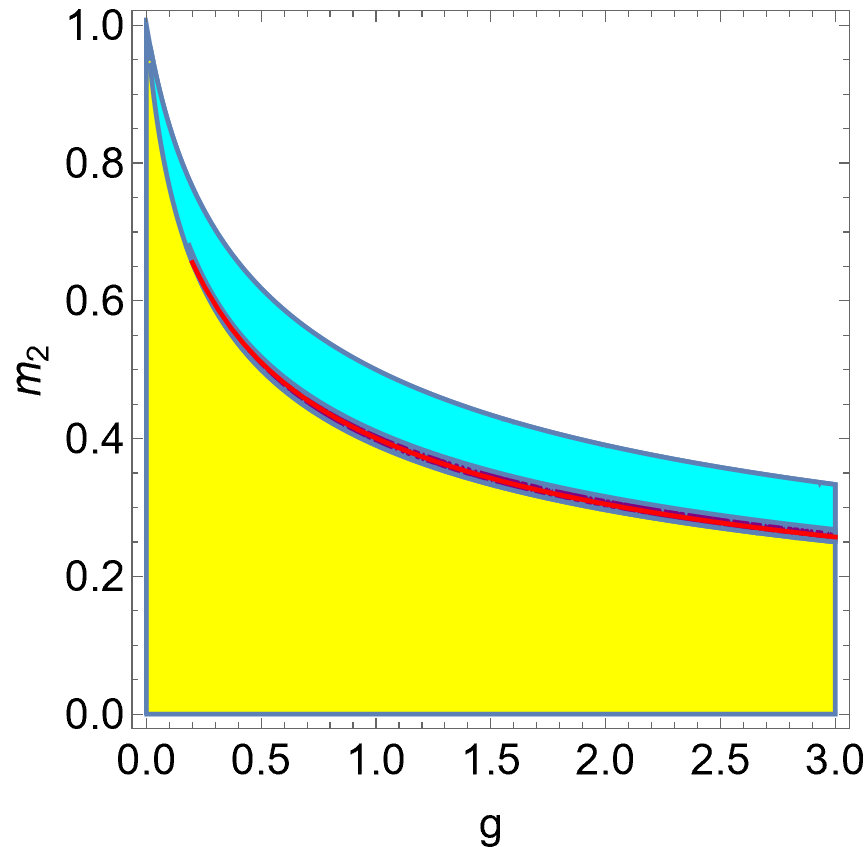}
    \caption{The bootstrapped solution compared with equation \eqref{eq:ansatz}, in red, for an estimate of $g_{c} =-0.0502729$. As in other plots, the curve and regions should extend to zero, but numerical error due to the removable singularity prevents this.}
    \label{fig:fitlambda4_fit}
\end{figure}

The curve is clearly convex for all submatrices of the Hankel matrix tested in $g\leq 0$, and all bootstrapped regions are well-approximated by a square root curve. For example, from Proposition \ref{prop:bound}, as we approach the point $g = -\frac{1}{16}$, the bound will behave as a square root singularity. For both the nine by nine and twenty-one by twenty-one submatrices of the Hankel matrix \eqref{eq:Hankel 2 matrix}, the curve can be well fit by a similarly structured curve of the form
\begin{equation}\label{eq:ansatz}
     -\frac{2 g_{c}(-1 + \sqrt{1 -  g/g_{c})}}{g},
\end{equation}
where $g_c$ is the conjectured critical point. This ansatz does an exceptionally good job estimating the solution for $g \geq 0$ as seen in Figure \ref{fig:fitlambda4_fit}.

It is tempting to conjecture that this ansatz is the actual form of $m_{2}$ since this would agree with the critical behaviour, curve fitting evidence, and the GUE limit as $g$ goes to zero, and it bears similarity to the solution of similar models in \cite{khalkhali2024coloured}. However, we can prove that the solution is not of this form using the results of Section \ref{sec:series expansion}. Observe that 
    \begin{equation*}
        -2 g_{c} \frac{-1 + \sqrt{1 - g/g_{c}}}{g} = 1 + 
        \frac{g}{2 g_{c}} + \frac{g^2}{8 g_{c}^2} + \frac{7 g^4}{128 g_{c}^4}+\frac{5 g^{3}}{64 g_{c}^3}+\mathcal{O}\left(g^5\right).
    \end{equation*}
    There is no value of $g_{c}$ that can make such a result consistent with equation \eqref{eq: m2_expansion}. We speculate that the actual solution to $m_{2}$ may not be an algebraic function and that it is just well-approximated by the above formula. All of this evidence leads us to conjecture the following.
\begin{conjecture}
The formal matrix model \ref{eq:2-matrix model} has  the asymptotic expansion of 
\begin{equation*}
      \frac{d}{d g}F_{0} \sim 2-2 \sqrt{1 - g/g_{c}} \quad g\rightarrow g_{c}
\end{equation*}
at a critical point $g_{c}$, which implies that the string susceptibility exponent is $\gamma =1/2$.

\end{conjecture}

\subsection{When $(g,\alpha,\beta) = (g,-g,g)$}
The search space of this model seems to be one. The bootstrap plots and series expansion of the moments at zero all appear to be the same as the $(g,g,g)$ configuration. This leads us to the following highly non-trivial conjecture.
\begin{conjecture}
    The  $(g,g,g)$ and  $(g,-g,g)$ 2-matrix models \eqref{eq:2-matrix model} have the same unique solution that satisfies the positivity constraints.
\end{conjecture}

Moreover, it would follow from the above conjecture that the series expansion at zero of $m_2$  for the $(g,-g,g)$ model as a function of $g$ is the same as in the $(g, g, g)$ model. Our argument for this is as follows. Solving the SDE gives the same equations for the moments in terms of $m_2$, except for a subset of equations whose sign is flipped compared to the $(g, g, g)$ case; however, one can observe that the latter can only happen for those moments whose $g \rightarrow 0$ limit is zero. This is because both models converge to the GUE case when $g\to 0$. 

Indeed, we can reason inductively. We observe that the moments are rational functions of $g$ and $m_2$, with the denominator for a word of length $2k$ being $(4g)^k$. Select a word of size $2k$ made up of an even number of A's and B's. Now, by induction, when we already know that 
$$m_2(0), \frac{d}{dg}m_2(0),\ldots, \frac{d^{k-1}}{dg^{k-1}}m_2(0)$$ agrees in the $(g, -g, g)$ case and in the $(g, g, g)$ case, we can look at the aforementioned moment whose $g \rightarrow 0$ limit is 0. 

When applying L'H\^{o}pital's rule $k$ times consecutively and substituting all previously known values for lower-order derivatives (which, by induction, are the same in both cases), this results in a linear equation in $\frac{d^{k}}{dg^{k}}m_2(0)$. This is the same equation in the $(g, -g, g)$ case as it is in the $(g, g, g)$ case, except that the sign of the right-hand side might be flipped. However, that can only happen when the right-hand side happens to be zero, as a result of our previous observation. Therefore, we must always obtain the same equation, which gives the same result for $\frac{d^{k}}{dg^{k}}m_2(0)$ in the $(g, -g, g)$ case and in the $(g, g, g)$ case.

A natural corollary of this observation is that all our reasoning about the critical behaviour of $m_2$ as a function of $g$ carries over from Section \ref{sec:ggg} into this case, and similarly, the critical exponent and asymptotic series expansion does so as well. We emphasize that there is no \textit{a priori} reason to believe that the above conjecture is true.

\subsection{When $(g,\alpha,\beta) = (g,g,-g)$}
The search space of this model seems to be two, relying on $m_{2}$ and $m_{4}$. One can see in Figure \ref{fig:bs_critical_-g_g_g} clear signs of a stable convergence to a solution for $g \geq 0$. The potential critical point in the given regions corresponds to the peak of the surface in the figure. As the Hankel matrix size increases to sixteen by sixteen, the convergence slows.

 \begin{figure}[H]
    \centering
    \includegraphics[width=0.7\textwidth]{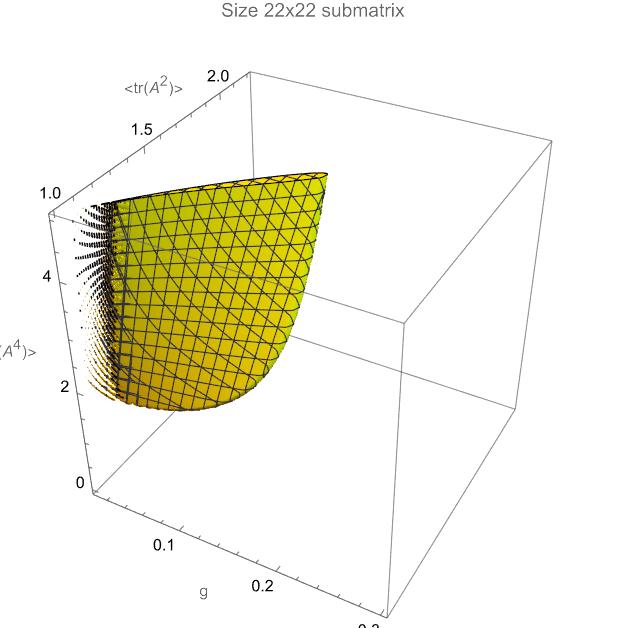}
     \caption{Bootstrapping results for the  $(g,\alpha,\beta) = (g,g,-g)$ 2-matrix model for a 22 by 22 submatrix of the positive semi-definite Hankel matrix. The stripped region near zero is the result of numerical error. Note this plot is actually a quite thin slice, which may be hard to discern from the presented fixed angle.}
    \label{fig:bs_critical_-g_g_g}
\end{figure}

\begin{figure}[H]
    \centering
     \subfigure{\includegraphics[width=0.45\textwidth]{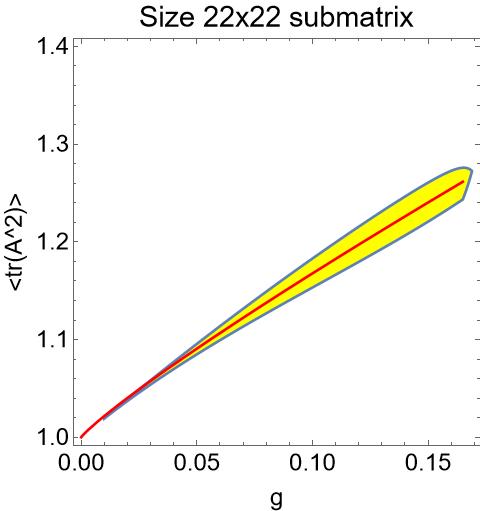}}
    \subfigure{\includegraphics[width=0.45\textwidth]{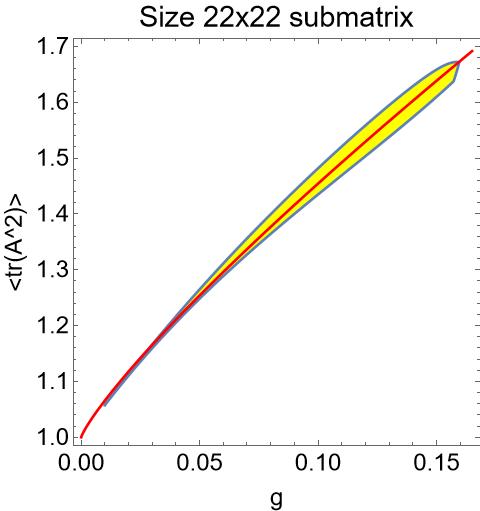}}
    \caption{Bootstrapping results for the  $(g,\alpha,\beta) = (g,g,-g)$ 2-matrix model on the level curves $m_{4} = 3$ on the left and $m_{4}=5$ on the right for a submatrix size 22 by 22. The red curves are the lines of best fit.}
    \label{fig:bs_critical_g_g_-g_fit}
\end{figure}

Ideally, we would hope for a finer curve; however, to deduce estimates of critical behaviour such a bootstrapped solution will suffice. Let us consider the level curves of the solution space for fixed $m_{4}$. Despite these level curves having different slopes, the power of the line of best fit remains approximately 0.85. For example, in Figure \ref{fig:bs_critical_g_g_-g_fit} the power is 0.886811 for the left subfigure when $m_{4} =3$ and 0.83865 for the right subfigure when $m_{4} = 3$. We can estimate there is a critical point for $m_{2}$ of this model near $g=0.165$ with a critical exponent of 0.85. This allows us to estimate the string susceptibility exponent since one can compute that 
 \begin{equation}
     \frac{d}{d g}F_{0} =  -\frac{1}{2} m_{4} - \frac{1}{2} m_{ 1,1,1,1}  +  m_{2,2}= \frac{m_{2}-1}{2g}.
 \end{equation}
 Hence, the estimated growth power of $\frac{d^2}{d g^2}F_{0}$ near the critical point is -0.15, making our estimate for the string susceptibility exponent $\gamma \approx 0.15$. This estimate is significantly distant from the usual two string susceptibility exponents, and may represent a new continuum limit.

The power series expansion of moments was computed to be 

\begin{equation*}
   m_{2} = 1+4g^2 +  96 g^4 + \mathcal{O}(g^{5}),
\end{equation*}
\begin{equation*}
   m_{4} = 2- g + 20 g^2 - 24 g^3 + \mathcal{O}(g^{4}),
\end{equation*}
 and the free energy can be found to be
 $$\lim_{N \rightarrow \infty}\frac{1}{N^2}\ln \mathcal{Z} = (F^{\text{GUE}}_{0})^{2}+ 2 g^2 + 12 g^4 + \mathcal{O}(g^{5}).$$

\subsection{When $(g,\alpha,\beta) = (-g,g,g)$}
This model is very similar to $(g,\alpha,\beta) = (g,g,-g)$ configuration.  The search space of this model also appears to be two, relying on $m_{2}$ and $m_{4}$. The bootstrap convergence also slows after a Hankel submatrix size of sixteen by sixteen is used.

   \begin{figure}[H]
    \centering
    \includegraphics[width=0.7\textwidth]{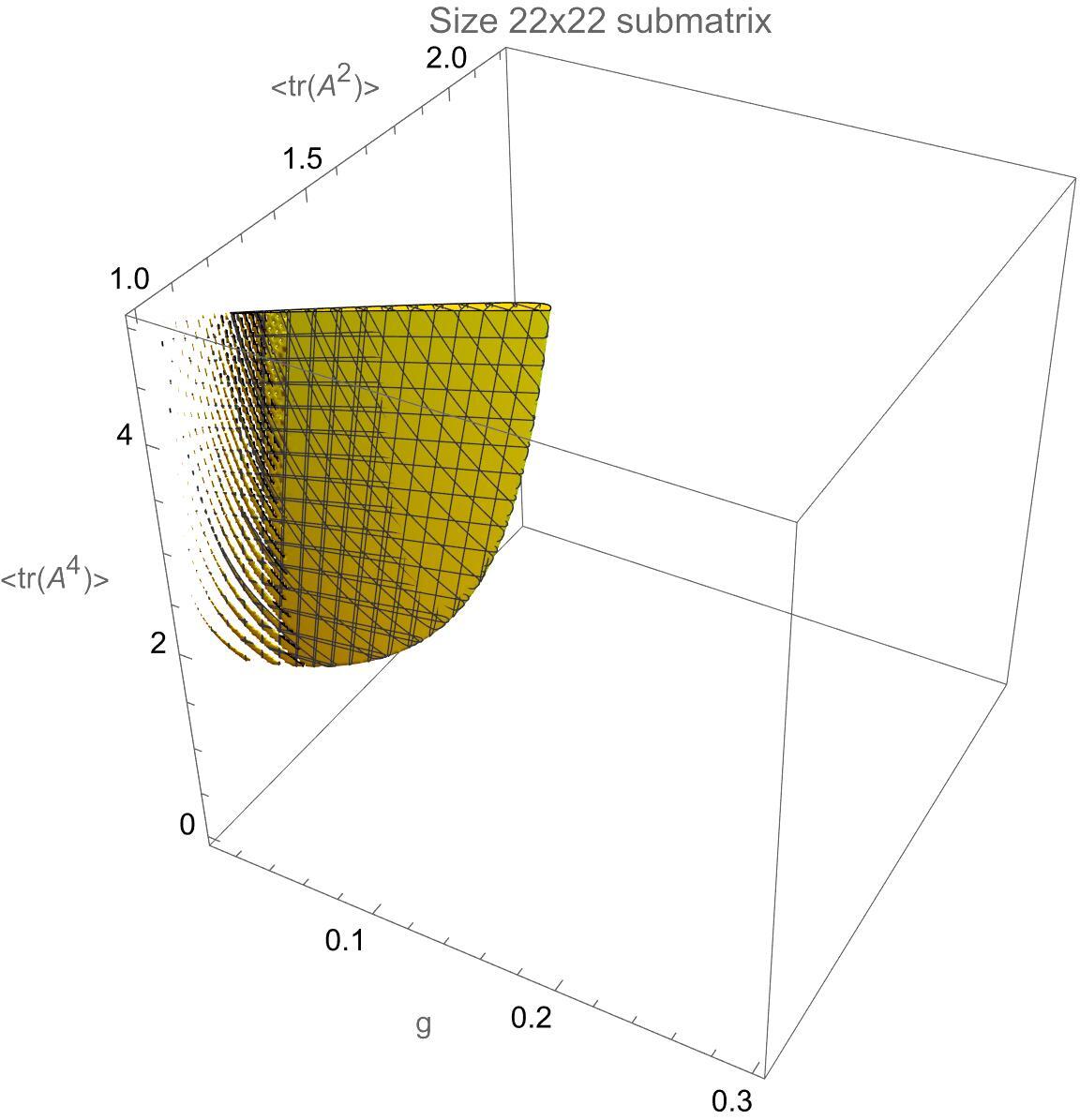}
    \caption{Bootstrapping results for the  $(g,\alpha,\beta) = (-g,g,g)$ 2-matrix model for various sizes of the submatrix of the positive semi-definite Hankel matrix. The stripped region near zero is the result of numerical error. Note this plot is actually a quite thin slice, which may be hard to discern from the presented fixed angle.}
    \label{fig:bs_critical_-g_g_g}
\end{figure}

\begin{figure}[H]
    \centering
     \subfigure{\includegraphics[width=0.45\textwidth]{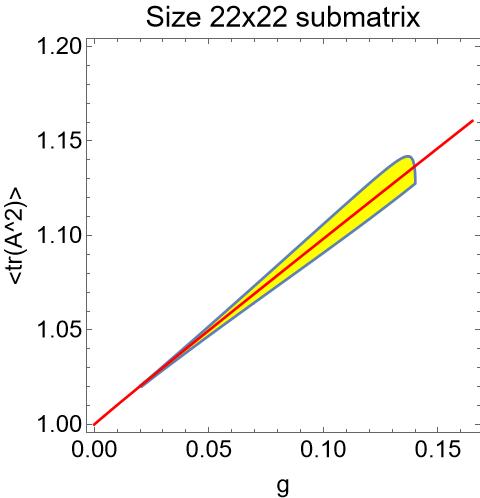}}
    \subfigure{\includegraphics[width=0.45\textwidth]{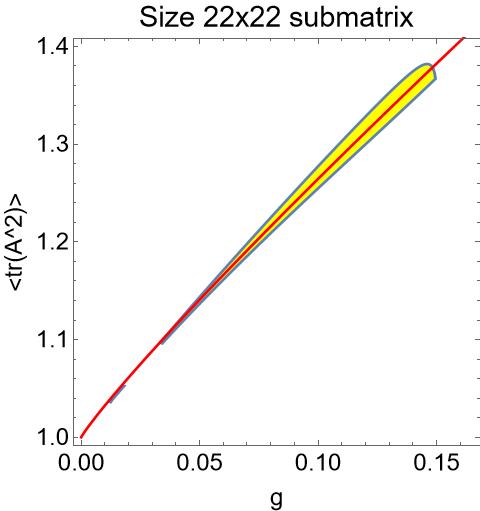}}
    \caption{Bootstrapping results for the  $(g,\alpha,\beta) = (g,g,-g)$ 2-matrix model on the level curves $m_{4} = 3$ on the left and $m_{4}=5$ on the right for a submatrix size 22 by 22. The red curves are the lines of best fit.}
    \label{fig:bs_critical_-g_g_g_fit}
\end{figure}

Just as in the previous section, we can also estimate the critical exponent of this model with level curves. In Figure \ref{fig:bs_critical_-g_g_g_fit} we plotted a line of best fit along the bootstrapped solution for level curves $m_{4}=3$ and $m_{4}=5$. The power of these curves was approximately 0.91 and 0.98, respectively. In general, it appears to be approximately 0.95. Just as in other models studied, we have that 
\begin{equation*}
    \frac{d}{d g} F_{0} = \frac{m_{2}-1}{2g},
\end{equation*}
so we estimate the string susceptibility exponent here to be $\gamma \approx 0.05$. Just as in the previous section, this may hint at a new continuum limit.

The power series expansion of moments was computed to be 
\begin{equation*}
   m_{2} = 1+ 4g^2 - 10 g^3 + 96 g^4 + \mathcal{O}(g^{5}),
\end{equation*}
and 
\begin{equation*}
   m_{4} = 2+g+10 g^2 - 24 g^3 + \mathcal{O}(g^{4}),
\end{equation*}
 and the free energy can be found to be
$$\lim_{N \rightarrow \infty}\frac{1}{N^2}\ln \mathcal{Z} =  (F^{\text{GUE}}_{0})^{2}+g + \frac{4}{3} g^3 + \frac{19}{2} g^4+ \mathcal{O}(g^{5}).$$

\section{The 3-matrix model}\label{sec:3matrix}
In this section we will study the following 3-matrix model
\begin{equation}\label{eq:3-matrix model}
	\mathcal{Z} = \int_{\mathcal{H}_{N}^{3}}\exp \left\{\frac{-N g}{3}\tr(A^3 + B^3 + C^3) - gN\tr (ABC + ACB) -\frac{N}{2}\tr (A^2 + B^2 + C^2)\right\}dA dB dC,
\end{equation}
which is a generalization of the three-colour model studied in \cite{eynard1998iterative,kostov2002exact}. As far as the authors can tell, this model is unsolved in the literature. For any word in $A,B$ and $C$ we denote the associated mixed moment in the large $N$ limit as 
\begin{equation*}
    m_{W} := \lim_{N \rightarrow \infty} \frac{1}{N}\langle \tr W\rangle.
\end{equation*}
The search space dimension of this model seems to be two and all moments used can generated through the SDE from $m_{A} :=m_{1}$ and $m_{AA} :=m_{2}$. 

Consider the lexicographically ordered tracial sequence $\{1, m_A, m_B,m_{C} , m_{AB}, m_{AC}, m_{BC} \ldots\}$. To obtain our bootstrap estimates we used submatrices of the following Hankel matrix:
\begin{equation*}\label{eq:Hankel 3 matrix}
\left[\begin{array}{ccccccc}
1 & m_{A} & m_{B}  & m_{C}  & m_{AB} & m_{AC} &\cdots\\
m_{A} & m_{A^2}  & m_{AB}  & m_{AC}  & m_{A^2 B}  & m_{A B C} & \cdots\\
m_{B} & m_{B A} & m_{B^2} & m_{B C}  & m_{B A B} & m_{BAC} & \cdots\\
m_{C}  & m_{C A} & m_{C B } & m_{C^2}  & m_{C A B}  & m_{C A C}  & \cdots \\
m_{B A} & m_{A^2 B} & m_{  B A  B} & m_{B A C} & m_{B A^2 B} & m_{B A^2 C } & \cdots\\
m_{ C A} & m_{ C A^2} & m_{ C A B} & m_{ C A C }  & m_{ C A^2 B}  & m_{C A^2 C} & \cdots\\
\vdots & \vdots & \vdots & \vdots & \vdots & \vdots &\ddots
\end{array}\right]\geq 0.
\end{equation*}

\begin{figure}[H]
    \centering
    \includegraphics[width=0.4\linewidth]{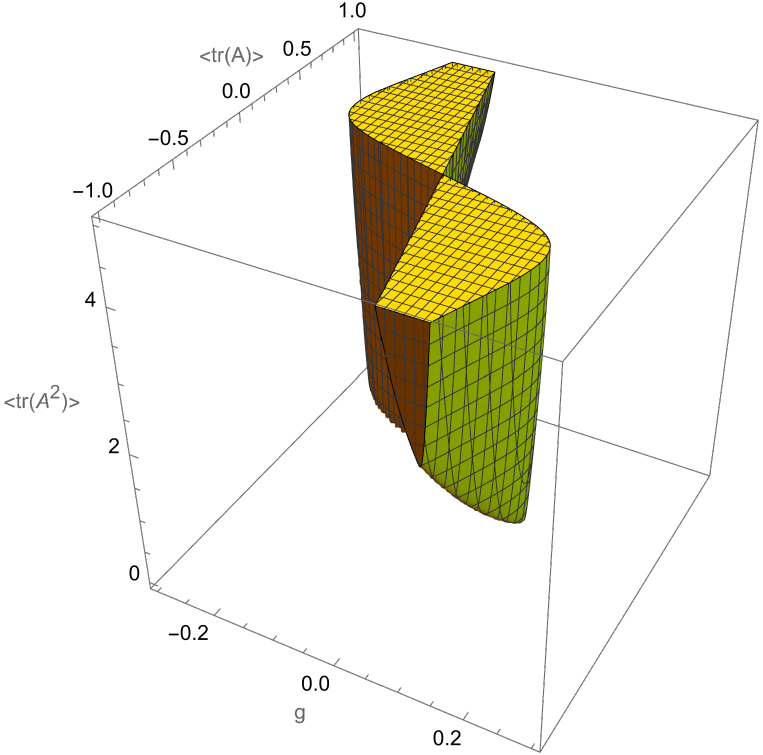}
    \includegraphics[width=0.4\linewidth]{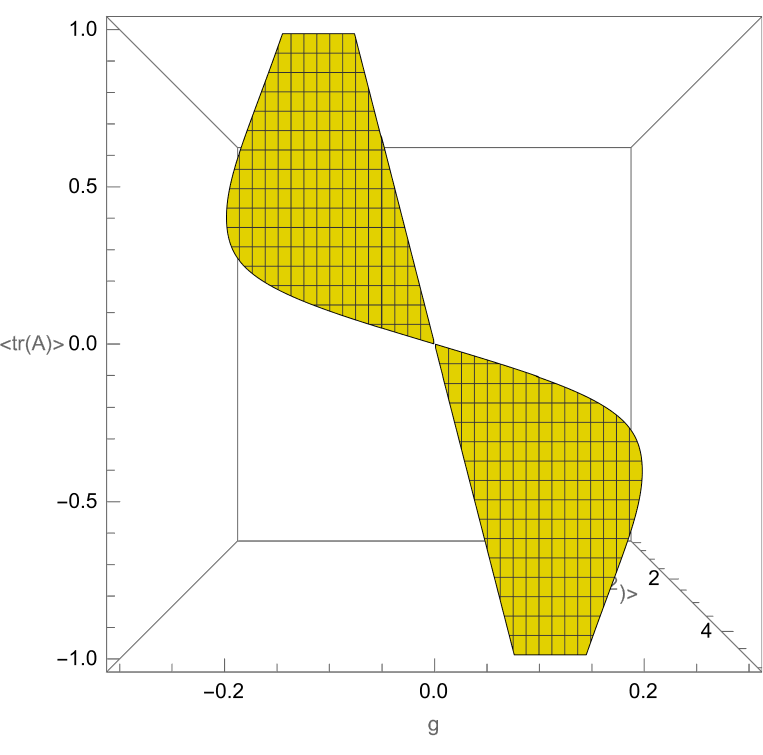}
    \caption{The bootstrapped estimate of the 3-matrix model \eqref{eq:3-matrix model} for a 14 by 14 submatrix of the Hankel matrix with a lexicographical basis.}
    \label{fig:bs_3matrix}
\end{figure}
The slices of $m_{1}$ for a fixed $m_{2}$ are seemingly asymmetric and are very reminiscent of the cubic model's solution for $m_{1}$ seen in Figure 7 of \cite{hessam2022noncommutative}. More submatrices are needed, however, to obtain a reasonable estimate of the location of a critical point and critical exponent.

Similarly, as in the case of the 2-matrix models studied in the previous section, we may write the free energy of the model with respect to the coupling constant and moments as 
	\begin{equation*}
		-\frac{d}{d g}\ln \mathcal{Z} = \frac{1}{3} \langle \tr (A^3 + B^3 + C^3) \rangle + \langle \tr ABC + \tr ACB \rangle. 
	\end{equation*}
In the large $N$ limit this will become 
\begin{equation*}
		-\frac{d}{d g}F_{0}= m_{3} + 2m_{ABC} = \frac{1-m_{2}}{g}.
	\end{equation*}

Applying the same process as in the previous section one can derive that
\begin{equation*}\label{eq: m2_der_expansion5}
   m_{2} = 1 -g  - 12 g^3  - 288 g^5 + \mathcal{O}(g^{7})
\end{equation*}
and
\begin{equation*}\label{eq: m4_der_expansion6}
   m_{AAAA} = 2+  6 g^2  + 114 g^4 + \mathcal{O}(g^{6}).
\end{equation*}
Hence, we have that 
\begin{equation*}
    \lim_{N \rightarrow \infty}\frac{1}{N^2}\ln \mathcal{Z} = (F_{0}^{\text{GUE}})^{3}-g - 4 g^3 - \frac{288}{5}g^{5} + \mathcal{O}(g^{7}).
\end{equation*}

\section{Conclusion and discussion}\label{Sec: conclusion}
In this paper, we investigated the critical phenomena of several multi-matrix models obtaining estimates for critical points and critical exponents via bootstrapping with positivity. In particular, for various 2-matrix models we find evidence of a string susceptibility exponent of $\gamma = 1/2$ placing those models potentially in the universality class of the Continuum Random Tree. For other 2-matrix models, we found evidence that $\gamma \approx 0.15$ and  $0.05$,  which are far from the critical exponents of any model known to the authors, certainly not the most typical $\gamma =1/2$ or $-1/2$. Such models may have a new continuum limit. This critical behaviour is of great interest in various approaches to quantum gravity such as 2d  conformal field theory, tensor models, and Dynamical Triangulations. It is the belief of the authors that the critical exponents of many unsolved matrix models can be accurately estimated by bootstrapping with positivity, the work done here serving as a first example. In theory, such bootstrapping techniques could also be applied to tensor models that can be transformed into matrix models \cite{lionni2018colored}, allowing for positivity constraints of the tensor models.

An additional feature of this work is that we were able to derive an elementary method of computing the series expansion of moments and the free energy of matrix models. This technique could be used in map enumeration problems that can be phrased as formal matrix integrals. In particular, it may allow for asymptotic growth estimates for the number of maps belonging to some class associated with a given model.

\subsection*{Acknowledgements}
We would like to thank the Fields Institute for organizing the Fields Undergraduate Summer Research Program in 2024. Andrei Parfeni and Brayden Smith were participants in this program while working on this paper. The project was proposed and
supervised by Masoud Khalkhali and Nathan Pagliaroli. We would also like to thank the Natural Sciences and Engineering Research Council of Canada (NSERC) for financial support.

\subsection*{Data availability} The code to reproduce the figures of this article will be shared upon request.

\subsection*{Conflict of interest}The authors have no competing interests to declare that pertain to the content of this article.

\printbibliography

@article{perez2024loop,
  title={The loop equations for noncommutative geometries on quivers},
  author={Perez-Sanchez, Carlos},
  journal={arXiv preprint arXiv:2409.03705},
  year={2024}
}

@article{schmudgen2020ten,
  title={Ten lectures on the moment problem},
  author={Schm{\"u}dgen, Konrad},
  journal={arXiv preprint arXiv:2008.12698},
  year={2020}
}

@article{kazakov1986ising,
  title={Ising model on a dynamical planar random lattice: exact solution},
  author={Kazakov, Vladimir A},
  journal={Physics Letters A},
  volume={119},
  number={3},
  pages={140--144},
  year={1986},
  publisher={Elsevier}
}

@article{hessam2022noncommutative,
  title={From noncommutative geometry to random matrix theory},
  author={Hessam, Hamed and Khalkhali, Masoud and Pagliaroli, Nathan and Verhoeven, Luuk S},
  journal={Journal of Physics A: Mathematical and Theoretical},
  volume={55},
  number={41},
  pages={413002},
  year={2022},
  publisher={IOP Publishing}
}

@article{tchoumakov2021bootstrapping,
  title={Bootstrapping bloch bands},
  author={Tchoumakov, Serguei and Florens, Serge},
  journal={Journal of Physics A: Mathematical and Theoretical},
  volume={55},
  number={1},
  pages={015203},
  year={2021},
  publisher={IOP Publishing}
}

@article{aikawa2022bootstrap,
  title={Bootstrap method in harmonic oscillator},
  author={Aikawa, Yu and Morita, Takeshi and Yoshimura, Kota},
  journal={Physics Letters B},
  volume={833},
  pages={137305},
  year={2022},
  publisher={Elsevier}
}

@article{berenstein2022bootstrapping,
  title={Bootstrapping more QM systems},
  author={Berenstein, David and Hulsey, George},
  journal={Journal of Physics A: Mathematical and Theoretical},
  volume={55},
  number={27},
  pages={275304},
  year={2022},
  publisher={IOP Publishing}
}

@article{bhattacharya2021numerical,
  title={Numerical bootstrap in quantum mechanics},
  author={Bhattacharya, Jyotirmoy and Das, Diptarka and Das, Sayan Kumar and Jha, Ankit Kumar and Kundu, Moulindu},
  journal={Physics Letters B},
  volume={823},
  pages={136785},
  year={2021},
  publisher={Elsevier}
}

@article{zeng2023feynman,
  title={Feynman integrals from positivity constraints},
  author={Zeng, Mao},
  journal={Journal of High Energy Physics},
  volume={2023},
  number={9},
  pages={1--43},
  year={2023},
  publisher={Springer}
}

@article{li2024analytic,
  title={Analytic trajectory bootstrap for matrix models},
  author={Li, Wenliang},
  journal={arXiv preprint arXiv:2407.08593},
  year={2024}
}

@article{chassaing2004random,
  title={Random planar lattices and integrated superBrownian excursion},
  author={Chassaing, Philippe and Schaeffer, Gilles},
  journal={Probability Theory and Related Fields},
  volume={128},
  pages={161--212},
  year={2004},
  publisher={Springer}
}

@article{angel2003growth,
  title={Growth and percolation on the uniform infinite planar triangulation},
  author={Angel, Omer},
  journal={Geometric And Functional Analysis},
  volume={13},
  pages={935--974},
  year={2003},
  publisher={Springer}
}

@article{le2013uniqueness,
  title={Uniqueness and universality of the Brownian map},
  author={Le Gall, Jean-Fran{\c{c}}ois},
  year={2013}
}

@article{marckert2006limit,
  title={Limit of normalized quadrangulations: the Brownian map},
  author={Marckert, Jean-Fran{\c{c}}ois and Mokkadem, Abdelkader},
  year={2006}
}

@inproceedings{le2010scaling,
  title={On the scaling limit of random planar maps with large faces},
  author={Le Gall, Jean-Fran{\c{c}}ois and Miermont, Gr{\'e}gory},
  booktitle={XVIth International Congress On Mathematical Physics: (With DVD-ROM)},
  pages={470--474},
  year={2010},
  organization={World Scientific}
}

@book{bleher2001random,
  title={Random matrix models and their applications},
  author={Bleher, Pavel and Its, Alexander},
  volume={40},
  year={2001},
  publisher={Cambridge university press}
}

@article{guhr1998random,
  title={Random-matrix theories in quantum physics: common concepts},
  author={Guhr, Thomas and M{\"u}ller--Groeling, Axel and Weidenm{\"u}ller, Hans A},
  journal={Physics Reports},
  volume={299},
  number={4-6},
  pages={189--425},
  year={1998},
  publisher={Elsevier}
}

@article{kazakov1999d,
  title={D-particles, matrix integrals and KP hierarchy},
  author={Kazakov, Vladimir A and Kostov, Ivan K and Nekrasov, Nikita},
  journal={Nuclear Physics B},
  volume={557},
  number={3},
  pages={413--442},
  year={1999},
  publisher={Elsevier}
}

@article{berenstein2009multi,
  title={Multi-matrix models and emergent geometry},
  author={Berenstein, David E and Hanada, Masanori and Hartnoll, Sean A},
  journal={Journal of High Energy Physics},
  volume={2009},
  number={02},
  pages={010},
  year={2009},
  publisher={IOP Publishing}
}

@article{kazakov1999two,
  title={Two-matrix model with ABAB interaction},
  author={Kazakov, Vladimir A and Zinn-Justin, Paul},
  journal={Nuclear Physics B},
  volume={546},
  number={3},
  pages={647--668},
  year={1999},
  publisher={Elsevier}
}

@article{zheng2023bootstrap,
  title={Bootstrap Method in Theoretical Physics},
  author={Zheng, Zechuan},
  journal={arXiv preprint arXiv:2401.00350},
  year={2023}
}

@article{burgdorf2012truncated,
  title={The truncated tracial moment problem},
  author={Burgdorf, Sabine and Klep, Igor},
  journal={Journal of Operator Theory},
  pages={141--163},
  year={2012},
  publisher={JSTOR}
}

@article{kostov2002exact,
  title={Exact solution of the three-color problem on a random lattice},
  author={Kostov, Ivan K},
  journal={Physics Letters B},
  volume={549},
  number={1-2},
  pages={245--252},
  year={2002},
  publisher={Elsevier}
}

@article{eynard1998iterative,
  title={An iterative solution of the three-colour problem on a random lattice},
  author={Eynard, B and Kristjansen, C},
  journal={Nuclear Physics B},
  volume={516},
  number={3},
  pages={529--542},
  year={1998},
  publisher={Elsevier}
}

@article{bergere2009universal,
  title={Universal scaling limits of matrix models, and (p, q) Liouville gravity},
  author={Bergere, Michel and Eynard, Bertrand},
  journal={arXiv preprint arXiv:0909.0854},
  year={2009}
}

@article{ambjorn2001lorentzian,
  title={Lorentzian 3d gravity with wormholes via matrix models},
  author={Ambj{\o}rn, Jan and Jurkiewicz, Jerzy and Loll, Renate and Vernizzi, Graziano},
  journal={Journal of High Energy Physics},
  volume={2001},
  number={09},
  pages={022},
  year={2001},
  publisher={IOP Publishing}
}

@article{bonzom2012random,
  title={Random tensor models in the large N limit: Uncoloring the colored tensor models},
  author={Bonzom, Valentin and Gurau, Razvan and Rivasseau, Vincent},
  journal={Physical Review D—Particles, Fields, Gravitation, and Cosmology},
  volume={85},
  number={8},
  pages={084037},
  year={2012},
  publisher={APS}
}

@book{lionni2018colored,
  title={Colored discrete spaces: higher dimensional combinatorial maps and quantum gravity},
  author={Lionni, Luca},
  year={2018},
  publisher={Springer}
}

@article{loll2019quantum,
  title={Quantum gravity from causal dynamical triangulations: a review},
  author={Loll, Renate},
  journal={Classical and Quantum Gravity},
  volume={37},
  number={1},
  pages={013002},
  year={2019},
  publisher={IOP Publishing}
}

@article{agishtein1992critical,
  title={Critical behavior of dynamically triangulated quantum gravity in four dimensions},
  author={Agishtein, Michael E and Migdal, Alexander A},
  journal={Nuclear Physics B},
  volume={385},
  number={1-2},
  pages={395--412},
  year={1992},
  publisher={Elsevier}
}

@book{pastur2011eigenvalue,
  title={Eigenvalue distribution of large random matrices},
  author={Pastur, Leonid Andreevich and Shcherbina, Mariya},
  number={171},
  year={2011},
  publisher={American Mathematical Soc.}
}

@article{eynard2019solutions,
  title={Solutions of loop equations are random matrices},
  author={Eynard, Bertrand},
  journal={arXiv preprint arXiv:1909.09372},
  year={2019}
}

@book{guionnet2019asymptotics,
  title={Asymptotics of random matrices and related models: the uses of Dyson-Schwinger equations},
  author={Guionnet, Alice},
  volume={130},
  year={2019},
  publisher={American Mathematical Soc.}
}

@incollection{budd2023lessons,
  title={Lessons from the mathematics of two-dimensional Euclidean quantum gravity},
  author={Budd, Timothy},
  booktitle={Handbook of Quantum Gravity},
  pages={1--55},
  year={2023},
  publisher={Springer}
}

@book{forrester2010log,
  title={Log-gases and random matrices (LMS-34)},
  author={Forrester, Peter J},
  year={2010},
  publisher={Princeton university press}
}

@book{deift2000orthogonal,
  title={Orthogonal Polynomials and Random Matrices: A Riemann-Hilbert Approach: A Riemann-Hilbert Approach},
  author={Deift, Percy},
  volume={3},
  year={2000},
  publisher={American Mathematical Soc.}
}

@article{johansson1998fluctuations,
  title={On fluctuations of eigenvalues of random Hermitian matrices},
  author={Johansson, Kurt},
  journal={Duke Mathematical Journal},
  volume={91},
  number={1},
  pages={151},
  year={1998},
  publisher={Duke University Press}
}

@article{ercolani2003asymptotics,
  title={Asymptotics of the partition function for random matrices via Riemann-Hilbert techniques and applications to graphical enumeration},
  author={Ercolani, Nicholas M and McLaughlin, KDT-R},
  journal={International Mathematics Research Notices},
  volume={2003},
  number={14},
  pages={755--820},
  year={2003},
  publisher={OUP}
}

@article{khalkhali2024coloured,
  title={Coloured combinatorial maps and quartic bi-tracial 2-matrix ensembles from noncommutative geometry},
  author={Khalkhali, Masoud and Pagliaroli, Nathan},
  journal={Journal of High Energy Physics},
  volume={2024},
  number={5},
  pages={1--28},
  year={2024},
  publisher={Springer}
}

@article{hessam2022bootstrapping,
  title={Bootstrapping dirac ensembles},
  author={Hessam, Hamed and Khalkhali, Masoud and Pagliaroli, Nathan},
  journal={Journal of Physics A: Mathematical and Theoretical},
  volume={55},
  number={33},
  pages={335204},
  year={2022},
  publisher={IOP Publishing}
}

@article{berenstein2023semidefinite,
  title={Semidefinite programming algorithm for the quantum mechanical bootstrap},
  author={Berenstein, David and Hulsey, George},
  journal={Physical Review E},
  volume={107},
  number={5},
  pages={L053301},
  year={2023},
  publisher={APS}
}

@article{cho2022bootstrapping,
  title={Bootstrapping the Ising model on the lattice},
  author={Cho, Minjae and Gabai, Barak and Lin, Ying-Hsuan and Rodriguez, Victor A and Sandor, Joshua and Yin, Xi},
  journal={arXiv preprint arXiv:2206.12538},
  year={2022}
}

@article{kazakov2023bootstrap,
  title={Bootstrap for lattice Yang-Mills theory},
  author={Kazakov, Vladimir and Zheng, Zechuan},
  journal={Physical Review D},
  volume={107},
  number={5},
  pages={L051501},
  year={2023},
  publisher={APS}
}

@article{kazakov2022analytic,
  title={Analytic and numerical bootstrap for one-matrix model and “unsolvable” two-matrix model},
  author={Kazakov, Vladimir and Zheng, Zechuan},
  journal={Journal of High Energy Physics},
  volume={2022},
  number={6},
  pages={1--61},
  year={2022},
  publisher={Springer}
}

@article{kazakov2024bootstrap,
  title={Bootstrap for Finite N Lattice Yang-Mills Theory},
  author={Kazakov, Vladimir and Zheng, Zechuan},
  journal={arXiv preprint arXiv:2404.16925},
  year={2024}
}

@article{guionnet2005combinatorial,
  title={Combinatorial aspects of matrix models},
  author={Guionnet, Alice and Maurel-Segala, Edouard},
  journal={arXiv preprint math/0503064},
  year={2005}
}

@article{guionnet2004first,
  title={First order asymptotics of matrix integrals; a rigorous approach towards the understanding of matrix models},
  author={Guionnet, Alice},
  journal={Communications in mathematical physics},
  volume={244},
  pages={527--569},
  year={2004},
  publisher={Springer}
}

@article{wigner1958distribution,
  title={On the distribution of the roots of certain symmetric matrices},
  author={Wigner, Eugene P},
  journal={Annals of Mathematics},
  volume={67},
  number={2},
  pages={325--327},
  year={1958},
  publisher={JSTOR}
}

@article{cates1985fractal,
  title={The fractal dimension and connectivity of random surfaces},
  author={Cates, ME},
  journal={Physics Letters B},
  volume={161},
  number={4-6},
  pages={363--367},
  year={1985},
  publisher={Elsevier}
}

@article{distler19902d,
  title={2D quantum gravity, topological field theory and the multicritical matrix models},
  author={Distler, Jacques},
  journal={Nuclear Physics B},
  volume={342},
  number={3},
  pages={523--538},
  year={1990},
  publisher={Elsevier}
}

@article{ambjorn2016generalized,
  title={Generalized multicritical one-matrix models},
  author={Ambj{\o}rn, Jan and Budd, Timothy and Makeenko, Yuri},
  journal={Nuclear Physics B},
  volume={913},
  pages={357--380},
  year={2016},
  publisher={Elsevier}
}

@article{kazakov1985critical,
  title={Critical properties of randomly triangulated planar random surfaces},
  author={Kazakov, Vo A and Kostov, IK and Migdal, AA},
  journal={Physics Letters B},
  volume={157},
  number={4},
  pages={295--300},
  year={1985},
  publisher={Elsevier}
}

@book{gross1991two,
  title={Two dimensional quantum gravity and random surfaces-8th Jerusalem winter school for theoretical physics},
  author={Gross, David J and Piran, Tsvi and Weinberg, Steven},
  volume={8},
  year={1991},
  publisher={World Scientific}
}

@incollection{t1993two,
  title={A two-dimensional model for mesons},
  author={'t Hooft, Gerard},
  booktitle={The Large N Expansion In Quantum Field Theory And Statistical Physics: From Spin Systems to 2-Dimensional Gravity},
  pages={94--103},
  year={1993},
  publisher={World Scientific}
}

@article{brezin1978planar,
  title={Planar diagrams},
  author={Br{\'e}zin, Edouard and Itzykson, Claude and Parisi, Giorgio and Zuber, Jean-Bernard},
  journal={Communications in Mathematical Physics},
  volume={59},
  pages={35--51},
  year={1978},
  publisher={Springer}
}

@article{kazakov2000solvable,
  title={Solvable matrix models},
  author={Kazakov, Vladimir A},
  journal={arXiv preprint hep-th/0003064},
  year={2000}
}

@incollection{eynard2011formal,
  title={Formal matrix integrals and combinatorics of maps},
  author={Eynard, Bertrand},
  booktitle={Random matrices, random processes and integrable systems},
  pages={415--442},
  year={2011},
  publisher={Springer}
}

@article{eynard2007invariants,
  title={Invariants of algebraic curves and topological expansion},
  author={Eynard, Bertrand and Orantin, Nicolas},
  journal={arXiv preprint math-ph/0702045},
  year={2007}
}

@article{eynard2016counting,
  title={Counting surfaces},
  author={Eynard, Bertrand and others},
  journal={Progress in Mathematical Physics},
  volume={70},
  pages={414},
  year={2016},
  publisher={Springer}
}

@article{kazakov1989appearance,
  title={The appearance of matter fields from quantum fluctuations of 2D-gravity},
  author={Kazakov, VA},
  journal={Modern Physics Letters A},
  volume={4},
  number={22},
  pages={2125--2139},
  year={1989},
  publisher={World Scientific}
}

@article{korchemsky1992matrix,
  title={Matrix model perturbed by higher order curvature terms},
  author={Korchemsky, Gregory P},
  journal={Modern Physics Letters A},
  volume={7},
  number={33},
  pages={3081--3100},
  year={1992},
  publisher={World Scientific}
}

@article{jonsson1998spectral,
  title={The spectral dimension of the branched polymer phase of two-dimensional quantum gravity},
  author={Jonsson, Thordur and Wheater, John F},
  journal={Nuclear Physics B},
  volume={515},
  number={3},
  pages={549--574},
  year={1998},
  publisher={Elsevier}
}

@article{aldous1991continuumI,
  title={The continuum random tree. I. An overview},
  author={Aldous, David},
  journal={The annals of probability 1991},
  volume={19},
  pages={1-28},
  year={1991},
  publisher={JSTOR}
}

@article{aldous1991continuumII,
  title={The continuum random tree. II. An overview},
  author={Aldous, David},
  journal={Stochastic analysis (Durham, 1990)},
  volume={167},
  pages={23--70},
  year={1991},
  publisher={Citeseer}
}

@article{aldous1993continuumIII,
  title={The continuum random tree III},
  author={Aldous, David},
  journal={The annals of probability},
  pages={248--289},
  year={1993},
  publisher={JSTOR}
}

@article{witten1990two,
  title={Two-dimensional gravity and intersection theory on moduli space},
  author={Witten, Edward},
  journal={Surveys in differential geometry},
  volume={1},
  number={1},
  pages={243--310},
  year={1990},
  publisher={International Press of Boston}
}

@book{lando2004graphs,
  title={Graphs on surfaces and their applications},
  author={Lando, Sergei K and Zvonkin, Alexander K and Zagier, Don Bernard},
  volume={141},
  year={2004},
  publisher={Springer}
}

@article{di19952d,
  title={2D gravity and random matrices},
  author={Di Francesco, Philippe and Ginsparg, Paul and Zinn-Justin, Jean},
  journal={Physics Reports},
  volume={254},
  number={1-2},
  pages={1--133},
  year={1995},
  publisher={Elsevier}
}

@book{Lando2004,
author="Lando, Sergei K.
and Zvonkin, Alexander K.",
title="Graphs on Surfaces and Their Applications",
year="2004",
publisher="Springer Berlin Heidelberg",
doi="10.1007/978-3-540-38361-1_1",
url="https://doi.org/10.1007/978-3-540-38361-1_1"
}

@Article{Lin2020,
author={Lin, Henry W.},
title={Bootstraps to strings: solving random matrix models with positivity},
journal={Journal of High Energy Physics},
year={2020},
month=jun,
day={15},
volume={2020},
number={6},
pages={90},
issn={1029-8479},
doi={10.1007/JHEP06(2020)090},
url={https://doi.org/10.1007/JHEP06(2020)090}
}

\appendix

\section{The 2-matrix model with $(g,\alpha,\beta) = (g,g,g)$ }

\subsection{The SDE}
The following are some examples of the SDE for this model:
\begin{align*}
A : 0 &= 1 - m_2 - g m_4  - 2 g m_{2,2} - g m_{1,1,1,1} \\ 
A^3 : 0 &= 2 m_2 - m_4  - g m_{6}  - 2 g m_{4,2}  - g m_{3,1,1, 1} \\ 
AB^2 : 0 &= m_2 - m_{2,2} - 2 g m_{4,2}  - g m_{3,1,1,1} - g m_{2,1,2,1} \\ 
BAB : 0 &= -m_{1,1,1,1} - 3 g m_{3,1,1,1} - g m_{2,1,2,1} \\ 
B^2A : 0 &= m_2 - m_{2,2} - 2 g m_{4,2}  - g m_{3,1,1,1} - g m_{2,1,2,1} \\ 
A^5 : 0 &= m_2^2 - 2 g m_{6,2} - g m_{5,1,1,1}  + 2 m_4  - m_{6}  - g m_{8} \\ 
A^3B^2 : 0 &= m_2^2 - g m_{6,2} - g m_{4,4} - g m_{3,1,1,3} - g m_{3,2,1,2} + m_{2,2} - m_{4,2}  \\ 
A^2BAB : 0 &= -g m_{5,1,1,1}  - g m_{3,1,1,3} - g m_{2,1,1,1,1,2} - g m_{2,1,1,2,1,1}  + m_{1,1,1,1} - m_{3,1,1,1} \\ 
A^2B^2A : 0 &= -g m_{6,2} - g m_{3,2,1,2} - g m_{2,1,1,1,1,2} - g m_{2,2,2,2} + 2 m_{2,2} - m_{4,2}  \\ 
ABA^2B^2 : 0 &= -g m_{4,1,2,1} - g m_{3,2,1,2} - g m_{2,1,1,1,1,2} - g m_{2,1,1,2,1,1}  + m_{2,2} - m_{2,1,2,1} \\ 
ABABA : 0 &= -g m_{5,1,1,1}  - 2 g m_{2,1,1,1,1,2} - g m_{1,1,1,1,1,1,1,1}  + 2 m_{1,1,1,1} - m_{3,1,1,1} \\ 
AB^2A^2 : 0 &= -g m_{6,2} - g m_{3,2,1,2} - g m_{2,1,1,1,1,2} - g m_{2,2,2,2} + 2 m_{2,2} - m_{4,2}  \\ 
AB^4 : 0 &= -g m_{6,2} - g m_{5,1,1,1}  - g m_{4,1,2,1} - g m_{4,4} + m_4  - m_{4,2}  \\ 
BA^3B : 0 &= -g m_{3,1,3,1} - 2 g m_{3,1,1,3} - g m_{3,2,1,2} - m_{3,1,1,1} \\ 
BA^2BA : 0 &= -g m_{4,1,2,1} - g m_{3,2,1,2} - g m_{2,1,1,1,1,2} - g m_{2,1,1,2,1,1}  + m_{2,2} - m_{2,1,2,1} \\ 
BABA^2 : 0 &= -g m_{5,1,1,1}  - g m_{3,1,1,3} - g m_{2,1,1,1,1,2} - g m_{2,1,1,2,1,1}  + m_{1,1,1,1} - m_{3,1,1,1} \\ 
BAB^3 : 0 &= -g m_{5,1,1,1}  - g m_{4,1,2,1} - g m_{3,1,3,1} - g m_{3,1,1,3} - m_{3,1,1,1} \\ 
B^2A^3 : 0 &= m_2^2 - g m_{6,2} - g m_{4,4} - g m_{3,1,1,3} - g m_{3,2,1,2} + m_{2,2} - m_{4,2}  \\ 
B^2AB^2 : 0 &= m_2^2 - 2 g m_{4,1,2,1} - g m_{3,1,3,1} - g m_{3,2,1,2} - m_{2,1,2,1} \\ 
B^3AB : 0 &= -g m_{5,1,1,1}  - g m_{4,1,2,1} - g m_{3,1,3,1} - g m_{3,1,1,3} - m_{3,1,1,1} 
\end{align*}

\subsection{The moments}\label{App:ggg_moments}
The following relations were found by solving some large set of SDE in Mathematica:
\begin{align*}
    m_{4} &= \frac{4 g m_{2}^2-m_{2}+1}{4 g}\\
    m_{2,2} &= \frac{1-m_{2}}{4 g}\\
    m_{1,1,1,1} &= \frac{-4 g m_{2}^2-m_{2}+1}{4 g}\\
    m_{6} &= \frac{12 g^2 m_{2}^3-12 g m_{2}^2+16 g m_{2}+m_{2}-1}{16 g^2}\\
    m_{4,2} &= \frac{-4 g^2 m_{2}^3-4 g m_{2}^2+8 g m_{2}+m_{2}-1}{16 g^2}\\
    m_{3,1,1,1} &= \frac{-4 g^2 m_{2}^3+4 g m_{2}^2+m_{2}-1}{16 g^2}\\
    m_{2,1,2,1} &= \frac{12 g^2 m_{2}^3+4 g m_{2}^2+m_{2}-1}{16 g^2}\\
    m_{8} &= \frac{16 g^3 m_{2}^4-80 g^2 m_{2}^3+148 g^2 m_{2}^2+24 g m_{2}^2-40 g m_{2}+12 g-m_{2}+1}{64 g^3}\\
    m_{6,2} &= \frac{-16 g^3 m_{2}^4+36 g^2 m_{2}^2+12 g m_{2}^2-24 g m_{2}+8 g-m_{2}+1}{64 g^3}\\
    m_{5,1,1,1} &= \frac{16 g^3 m_{2}^4+32 g^2 m_{2}^3-28 g^2 m_{2}^2-8 g m_{2}+4 g-m_{2}+1}{64 g^3}\\
    m_{4,1,2,1} &= \frac{16 g^3 m_{2}^4-16 g^2 m_{2}^3+20 g^2 m_{2}^2-4 g m_{2}^2-m_{2}+1}{64 g^3}\\
    m_{4,4} &= \frac{-16 g^3 m_{2}^4+36 g^2 m_{2}^2+8 g m_{2}^2-16 g m_{2}+4 g-m_{2}+1}{64 g^3}\\
    m_{3,1,3,1} &= \frac{-48 g^3 m_{2}^4-16 g^2 m_{2}^3+20 g^2 m_{2}^2-8 g m_{2}^2+8 g m_{2}-4 g-m_{2}+1}{64 g^3}\\
    m_{3,1,1,3} &=\frac{16 g^3 m_{2}^4+16 g^2 m_{2}^3-12 g^2 m_{2}^2-4 g m_{2}^2-m_{2}+1}{64 g^3} \\
    m_{3,2,1,2} &= \frac{16 g^3 m_{2}^4+4 g^2 m_{2}^2-8 g m_{2}+4 g-m_{2}+1}{64 g^3}\\
    m_{2,1,1,1,1,2} &= \frac{-16 g^3 m_{2}^4-28 g^2 m_{2}^2-4 g m_{2}^2-8 g m_{2}+8 g-m_{2}+1}{64 g^3}\\
    m_{2,1,1,2,1,1} &= \frac{-16 g^3 m_{2}^4-32 g^2 m_{2}^3+4 g^2 m_{2}^2-8 g m_{2}^2+4 g-m_{2}+1}{64 g^3}\\
    m_{2,2,2,2} &= \frac{16 g^3 m_{2}^4+16 g^2 m_{2}^3-12 g^2 m_{2}^2+8 g m_{2}^2-24 g m_{2}+12 g-m_{2}+1}{64 g^3}\\
    m_{1,1,1,1,1,1,1,1} &= \frac{16 g^3 m_{2}^4-16 g^2 m_{2}^3-44 g^2 m_{2}^2-8 g m_{2}^2-8 g m_{2}+12 g-m_{2}+1}{64 g^3}\\
\end{align*}

\section{The 2-matrix model with $(g,\alpha,\beta) = (g,-g,g)$ }

\subsection{The SDE}
The following are some examples of the SDE for this model:
\begin{align*}
A : 0 &= 1 - m_2 - g m_4  - 2 g m_{2,2} + g m_{1,1,1,1} \\ 
A^3 : 0 &= 2 m_2 - m_4  - g m_{6}  - 2 g m_{4,2}  + g m_{3,1,1,1} \\ 
AB^2 : 0 &= m_2 - m_{2,2} - 2 g m_{4,2}  + g m_{3,1,1,1} - g m_{2,1,2,1} \\ 
BAB : 0 &= -m_{1,1,1,1} - 3 g m_{3,1,1,1} + g m_{2,1,2,1} \\ 
B^2A : 0 &= m_2 - m_{2,2} - 2 g m_{4,2}  + g m_{3,1,1,1} - g m_{2,1,2,1} \\ 
A^5 : 0 &= m_2^2 - 2 g m_{6,2} + g m_{5,1,1,1}  + 2 m_4  - m_{6}  - g m_{8} \\ 
A^3B^2 : 0 &= m_2^2 - g m_{6,2} - g m_{4,4} + g m_{3,1,1,3} - g m_{3,2,1,2} + m_{2,2} - m_{4,2}  \\ 
A^2BAB : 0 &= -g m_{5,1,1,1}  - g m_{3,1,1,3} - g m_{2,1,1,1,1,2} + g m_{2,1,1,2,1,1}  + m_{1,1,1,1} - m_{3,1,1,1} \\ 
A^2B^2A : 0 &= -g m_{6,2} - g m_{3,2,1,2} + g m_{2,1,1,1,1,2} - g m_{2,2,2,2} + 2 m_{2,2} - m_{4,2}  \\ 
ABA^2B^2 : 0 &= -g m_{4,1,2,1} - g m_{3,2,1,2} + g m_{2,1,1,1,1,2} - g m_{2,1,1,2,1,1}  + m_{2,2} - m_{2,1,2,1} \\ 
ABABA : 0 &= -g m_{5,1,1,1}  - 2 g m_{2,1,1,1,1,2} + g m_{1,1,1,1,1,1,1,1}  + 2 m_{1,1,1,1} - m_{3,1,1,1} \\ 
AB^2A^2 : 0 &= -g m_{6,2} - g m_{3,2,1,2} + g m_{2,1,1,1,1,2} - g m_{2,2,2,2} + 2 m_{2,2} - m_{4,2}  \\ 
AB^4 : 0 &= -g m_{6,2} + g m_{5,1,1,1}  - g m_{4,1,2,1} - g m_{4,4} + m_4  - m_{4,2}  \\ 
BA^3B : 0 &= -g m_{3,1,3,1} - 2 g m_{3,1,1,3} + g m_{3,2,1,2} - m_{3,1,1,1} \\ 
BA^2BA : 0 &= -g m_{4,1,2,1} - g m_{3,2,1,2} + g m_{2,1,1,1,1,2} - g m_{2,1,1,2,1,1}  + m_{2,2} - m_{2,1,2,1} \\ 
BABA^2 : 0 &= -g m_{5,1,1,1}  - g m_{3,1,1,3} - g m_{2,1,1,1,1,2} + g m_{2,1,1,2,1,1}  + m_{1,1,1,1} - m_{3,1,1,1} \\ 
BAB^3 : 0 &= -g m_{5,1,1,1}  + g m_{4,1,2,1} - g m_{3,1,3,1} - g m_{3,1,1,3} - m_{3,1,1,1} \\ 
B^2A^3 : 0 &= m_2^2 - g m_{6,2} - g m_{4,4} + g m_{3,1,1,3} - g m_{3,2,1,2} + m_{2,2} - m_{4,2}  \\ 
B^2AB^2 : 0 &= m_2^2 - 2 g m_{4,1,2,1} + g m_{3,1,3,1} - g m_{3,2,1,2} - m_{2,1,2,1} \\ 
B^3AB : 0 &= -g m_{5,1,1,1}  + g m_{4,1,2,1} - g m_{3,1,3,1} - g m_{3,1,1,3} - m_{3,1,1,1} 
\end{align*}

\subsection{The moments}

The following relations were found by solving some large set of SDE in Mathematica:
\begin{align*}
    m_{4} &= \frac{4 g m_{2}^2-m_{2}+1}{4 g}\\
    m_{2,2} &= \frac{1-m_{2}}{4 g}\\
    m_{1,1,1,1} &= \frac{4 g m_{2}^2+m_{2}-1}{4 g}\\
    m_{6} &= \frac{12 g^2 m_{2}^3-12 g m_{2}^2+16 g m_{2}+m_{2}-1}{16 g^2}\\
    m_{4,2} &= \frac{-4 g^2 m_{2}^3-4 g m_{2}^2+8 g m_{2}+m_{2}-1}{16 g^2}\\
    m_{3,1,1,1} &= \frac{-4 g^2 m_{2}^3+4 g m_{2}^2+m_{2}-1}{16 g^2}\\
    m_{2,1,2,1} &= \frac{12 g^2 m_{2}^3+4 g m_{2}^2+m_{2}-1}{16 g^2}\\
    m_{8} &= \frac{16 g^3 m_{2}^4-80 g^2 m_{2}^3+148 g^2 m_{2}^2+24 g m_{2}^2-40 g m_{2}+12 g-m_{2}+1}{64 g^3}\\
    m_{6,2} &= \frac{-16 g^3 m_{2}^4+36 g^2 m_{2}^2+12 g m_{2}^2-24 g m_{2}+8 g-m_{2}+1}{64 g^3}\\
    m_{5,1,1,1} &= \frac{16 g^3 m_{2}^4+32 g^2 m_{2}^3-28 g^2 m_{2}^2-8 g m_{2}+4 g-m_{2}+1}{64 g^3}\\
    m_{4,1,2,1} &= \frac{16 g^3 m_{2}^4-16 g^2 m_{2}^3+20 g^2 m_{2}^2-4 g m_{2}^2-m_{2}+1}{64 g^3}\\
    m_{4,4} &= \frac{-16 g^3 m_{2}^4+36 g^2 m_{2}^2+8 g m_{2}^2-16 g m_{2}+4 g-m_{2}+1}{64 g^3}\\
    m_{3,1,3,1} &= \frac{-48 g^3 m_{2}^4-16 g^2 m_{2}^3+20 g^2 m_{2}^2-8 g m_{2}^2+8 g m_{2}-4 g-m_{2}+1}{64 g^3}\\
    m_{3,1,1,3} &=\frac{16 g^3 m_{2}^4+16 g^2 m_{2}^3-12 g^2 m_{2}^2-4 g m_{2}^2-m_{2}+1}{64 g^3} \\
    m_{3,2,1,2} &= \frac{16 g^3 m_{2}^4+4 g^2 m_{2}^2-8 g m_{2}+4 g-m_{2}+1}{64 g^3}\\
    m_{2,1,1,1,1,2} &= \frac{-16 g^3 m_{2}^4-28 g^2 m_{2}^2-4 g m_{2}^2-8 g m_{2}+8 g-m_{2}+1}{64 g^3}\\
    m_{2,1,1,2,1,1} &= \frac{-16 g^3 m_{2}^4-32 g^2 m_{2}^3+4 g^2 m_{2}^2-8 g m_{2}^2+4 g-m_{2}+1}{64 g^3}\\
    m_{2,2,2,2} &= \frac{16 g^3 m_{2}^4+16 g^2 m_{2}^3-12 g^2 m_{2}^2+8 g m_{2}^2-24 g m_{2}+12 g-m_{2}+1}{64 g^3}\\
    m_{1,1,1,1,1,1,1,1} &= \frac{16 g^3 m_{2}^4-16 g^2 m_{2}^3-44 g^2 m_{2}^2-8 g m_{2}^2-8 g m_{2}+12 g-m_{2}+1}{64 g^3}\\
\end{align*}

For words of length at most 8, these equations are identical to the ones given by the $(g, g, g)$ 2-matrix model, except for the sign of $m_{ABAB}$ being flipped.

\section{The 2-matrix model with $(g,\alpha,\beta) = (g,g,-g)$ }

\subsection{The SDE}
The following are some examples of the SDE for this model:
\begin{align*}
A : 0 &= 1 - m_2 - g m_4  + 2 g m_{2,2} - g m_{1,1,1,1} \\ 
A^3 : 0 &= 2 m_2 - m_4  - g m_{6}  + 2 g m_{4,2}  - g m_{3,1,1,1} \\ 
AB^2 : 0 &= m_2 - m_{2,2} - g m_{3,1,1,1} + g m_{2,1,2,1} \\ 
BAB : 0 &= -m_{1,1,1,1} + g m_{3,1,1,1} - g m_{2,1,2,1} \\ 
B^2A : 0 &= m_2 - m_{2,2} - g m_{3,1,1,1} + g m_{2,1,2,1} \\ 
A^5 : 0 &= m_2^2 + 2 g m_{6,2} - g m_{5,1,1,1}  + 2 m_4  - m_{6}  - g m_{8} \\ 
A^3B^2 : 0 &= m_2^2 - g m_{6,2} + g m_{4,4} - g m_{3,1,1,3} + g m_{3,2,1,2} + m_{2,2} - m_{4,2}  \\ 
A^2BAB : 0 &= -g m_{5,1,1,1}  + g m_{3,1,1,3} + g m_{2,1,1,1,1,2} - g m_{2,1,1,2,1,1}  + m_{1,1,1,1} - m_{3,1,1,1} \\ 
A^2B^2A : 0 &= -g m_{6,2} + g m_{3,2,1,2} - g m_{2,1,1,1,1,2} + g m_{2,2,2,2} + 2 m_{2,2} - m_{4,2}  \\ 
ABA^2B^2 : 0 &= -g m_{4,1,2,1} + g m_{3,2,1,2} - g m_{2,1,1,1,1,2} + g m_{2,1,1,2,1,1}  + m_{2,2} - m_{2,1,2,1} \\ 
ABABA : 0 &= -g m_{5,1,1,1}  + 2 g m_{2,1,1,1,1,2} - g m_{1,1,1,1,1,1,1,1}  + 2 m_{1,1,1,1} - m_{3,1,1,1} \\ 
AB^2A^2 : 0 &= -g m_{6,2} + g m_{3,2,1,2} - g m_{2,1,1,1,1,2} + g m_{2,2,2,2} + 2 m_{2,2} - m_{4,2}  \\ 
AB^4 : 0 &= g m_{6,2} - g m_{5,1,1,1}  + g m_{4,1,2,1} - g m_{4,4} + m_4  - m_{4,2}  \\ 
BA^3B : 0 &= -g m_{3,1,3,1} + 2 g m_{3,1,1,3} - g m_{3,2,1,2} - m_{3,1,1,1} \\ 
BA^2BA : 0 &= -g m_{4,1,2,1} + g m_{3,2,1,2} - g m_{2,1,1,1,1,2} + g m_{2,1,1,2,1,1}  + m_{2,2} - m_{2,1,2,1} \\ 
BABA^2 : 0 &= -g m_{5,1,1,1}  + g m_{3,1,1,3} + g m_{2,1,1,1,1,2} - g m_{2,1,1,2,1,1}  + m_{1,1,1,1} - m_{3,1,1,1} \\ 
BAB^3 : 0 &= g m_{5,1,1,1}  - g m_{4,1,2,1} + g m_{3,1,3,1} - g m_{3,1,1,3} - m_{3,1,1,1} \\ 
B^2A^3 : 0 &= m_2^2 - g m_{6,2} + g m_{4,4} - g m_{3,1,1,3} + g m_{3,2,1,2} + m_{2,2} - m_{4,2}  \\ 
B^2AB^2 : 0 &= m_2^2 + 2 g m_{4,1,2,1} - g m_{3,1,3,1} - g m_{3,2,1,2} - m_{2,1,2,1} \\ 
B^3AB : 0 &= g m_{5,1,1,1}  - g m_{4,1,2,1} + g m_{3,1,3,1} - g m_{3,1,1,3} - m_{3,1,1,1}
\end{align*}
\subsection{The moments}
The following relations were found by solving some large set of SDE in Mathematica:
\begin{align*}
    m_{2,2}&= \frac{-1 + m_{2} + g m_{2} + g m_{4}}{3 g}\\ 
m_{1,1,1,1} &= -\frac{-1 + m_{2} - 2 g m_{2} + g m_{4}}{3 g)}\\ 
m_{6} &= \frac{-6 - 4 g + 6 m_{2} + 16 g m_{2} - 5 g^2 m_{2} + 3 g^2 m_{2}^2 - 
    6 g m_{4} + 16 g^2 m_{4}}{12 g^2}\\ 
m_{4,2} &= \frac{-2 - 2 g + 2 m_{2} - 2 g m_{2} - g^2 m_{2} + 3 g^2 m_{2}^2 + 2 g m_{4} + 
    8 g^2 m_{4}}{12 g^2}\\
m_{3,1,1,1} &= \frac{2 - 2 m_{2} + 4 g m_{2} + 3 g^2 m_{2} + 3 g^2 m_{2}^2 - 
    2 g m_{4}}{12 g^2}\\ 
m_{2,1,2,1}&= \frac{-2 + 2 m_{2} - 4 g m_{2} + 3 g^2 m_{2} + 3 g^2 m_{2}^2 + 
    2 g m_{4}}{12 g^2}\\
m_{8} &= \frac{1}{48 g^3}(12 - 16 g - 15 g^2 - 12 m_{2} - 56 g m_{2} + 18 g^2 m_{2} - 
    24 g^3 m_{2} + 49 g^2 m_{2}^2 - 8 g^3 m_{2}^2 + 36 g m_{4} + 16 g^2 m_{4}\\
    &+ 
    60 g^3 m_{4} + 52 g^3 m_{2} m_{4})\\
\end{align*}

\section{The 2-matrix model with $(g,\alpha,\beta) = (-g,g,g)$ }

\subsection{The SDE}\label{App: 2-matrix SDE}
The following are some examples of the SDE for this model:
\begin{align*}
A : 0 &=  1 - m_2 + g m_4  - 2 g m_{2,2} - g m_{1,1,1,1} \\ 
A^3 : 0 &= 2 m_2 - m_4  + g m_{6}  - 2 g m_{4,2}  - g m_{3,1,1,1} \\ 
AB^2 : 0 &= m_2 - m_{2,2} - g m_{3,1,1,1} - g m_{2,1,2,1} \\ 
BAB : 0 &= -m_{1,1,1,1} - g m_{3,1,1,1} - g m_{2,1,2,1} \\ 
B^2A : 0 &= m_2 - m_{2,2} - g m_{3,1,1,1} - g m_{2,1,2,1} \\ 
A^5 : 0 &= m_2^2 - 2 g m_{6,2} - g m_{5,1,1,1}  + 2 m_4  - m_{6}  + g m_{8} \\ 
A^3B^2 : 0 &= m_2^2 + g m_{6,2} - g m_{4,4} - g m_{3,1,1,3} - g m_{3,2,1,2} + m_{2,2} - m_{4,2}  \\ 
A^2BAB : 0 &= g m_{5,1,1,1}  - g m_{3,1,1,3} - g m_{2,1,1,1,1,2} - g m_{2,1,1,2,1,1}  + m_{1,1,1,1} - m_{3,1,1,1} \\ 
A^2B^2A : 0 &= g m_{6,2} - g m_{3,2,1,2} - g m_{2,1,1,1,1,2} - g m_{2,2,2,2} + 2 m_{2,2} - m_{4,2}  \\ 
ABA^2B^2 : 0 &= g m_{4,1,2,1} - g m_{3,2,1,2} - g m_{2,1,1,1,1,2} - g m_{2,1,1,2,1,1}  + m_{2,2} - m_{2,1,2,1} \\ 
ABABA : 0 &= g m_{5,1,1,1}  - 2 g m_{2,1,1,1,1,2} - g m_{1,1,1,1,1,1,1,1}  + 2 m_{1,1,1,1} - m_{3,1,1,1} \\ 
AB^2A^2 : 0 &= g m_{6,2} - g m_{3,2,1,2} - g m_{2,1,1,1,1,2} - g m_{2,2,2,2} + 2 m_{2,2} - m_{4,2}  \\ 
AB^4 : 0 &= -g m_{6,2} - g m_{5,1,1,1}  - g m_{4,1,2,1} + g m_{4,4} + m_4  - m_{4,2}  \\ 
BA^3B : 0 &= g m_{3,1,3,1} - 2 g m_{3,1,1,3} - g m_{3,2,1,2} - m_{3,1,1,1} \\ 
BA^2BA : 0 &= g m_{4,1,2,1} - g m_{3,2,1,2} - g m_{2,1,1,1,1,2} - g m_{2,1,1,2,1,1}  + m_{2,2} - m_{2,1,2,1} \\ 
BABA^2 : 0 &= g m_{5,1,1,1}  - g m_{3,1,1,3} - g m_{2,1,1,1,1,2} - g m_{2,1,1,2,1,1}  + m_{1,1,1,1} - m_{3,1,1,1} \\ 
BAB^3 : 0 &= -g m_{5,1,1,1}  - g m_{4,1,2,1} - g m_{3,1,3,1} + g m_{3,1,1,3} - m_{3,1,1,1} \\ 
B^2A^3 : 0 &= m_2^2 + g m_{6,2} - g m_{4,4} - g m_{3,1,1,3} - g m_{3,2,1,2} + m_{2,2} - m_{4,2}  \\ 
B^2AB^2 : 0 &= m_2^2 - 2 g m_{4,1,2,1} - g m_{3,1,3,1} + g m_{3,2,1,2} - m_{2,1,2,1} \\ 
B^3AB : 0 &= -g m_{5,1,1,1}  - g m_{4,1,2,1} - g m_{3,1,3,1} + g m_{3,1,1,3} - m_{3,1,1,1} 
\end{align*}

\subsection{The moments}
The following relations were found by solving some large set of SDE in Mathematica:
\begin{align*}
    m_{2,2} &= \frac{1 - m_{2} + g m_{2} + g m_{4}}{3 g}\\ 
m_{1,1,1,1}&= \frac{1 - m_{2} - 2 g m_{2} + g m_{4}}{3 g}\\
m_{6} &= \frac{-6 + 4 g + 6 m_{2} - 16 g m_{2} - 5 g^2 m_{2} + 3 g^2 m_{2}^2 + 
    6 g m_{4} + 16 g^2 m_{4}}{12 g^2}\\ 
m_{4,2} &= \frac{-2 + 2 g + 2 m_{2} + 2 g m_{2} - g^2 m_{2} + 3 g^2 m_{2}^2 - 2 g m_{4} + 
    8 g^2 m_{4}}{12 g^2}\\
m_{3,1,1,1} &= \frac{-2 + 2 m_{2} + 4 g m_{2} - 3 g^2 m_{2} - 3 g^2 m_{2}^2 - 
    2 g m_{4}}{12 g^2}\\ 
m_{2,1,2,1} &= \frac{-2 + 2 m_{2} + 4 g m_{2} + 3 g^2 m_{2} + 3 g^2 m_{2}^2 - 
    2 g m_{4}}{12 g^2}\\ 
m_{8}&= \frac{1}{48 g^3}(-12 - 16 g + 15 g^2 + 12 m_{2} - 56 g m_{2} - 18 g^2 m_{2} - 
    24 g^3 m_{2} - 49 g^2 m_{2}^2 - 8 g^3 m_{2}^2 \\
    &+ 36 g m_{4} - 16 g^2 m_{4} + 
    60 g^3 m_{4} + 52 g^3 m_{2} m_{4})\\ 
\end{align*}
\section{The 3-matrix model}
\subsection{The SDE}
The following are some examples of the SDE for this model:
\begin{align*}
A : 0 &= 1 - m_{AA} - g m_{AAA} - 2 g m_{ABC} \\ 
B : 0 &= -m_{AB} - 3 g m_{AAB} \\ 
C : 0 &= -m_{AB} - 3 g m_{AAB} \\ 
A^2 : 0 &= 2 m_{A} - 2 g m_{AABC} - m_{AAA} - g m_{AAAA} \\ 
AB : 0 &= m_{A} - g m_{AABC} - g m_{ABAC} - m_{AAB} - g m_{AAAB} \\ 
AC : 0 &= m_{A} - g m_{AABC} - g m_{ABAC} - m_{AAB} - g m_{AAAB} \\ 
BA : 0 &= m_{A} - g m_{AABC} - g m_{ABAC} - m_{AAB} - g m_{AAAB} \\ 
B^2 : 0 &= -m_{AAB} - 2 g m_{AAAB} - g m_{AABB} \\ 
BC : 0 &= -g m_{AABC} - g m_{ABAB} - m_{ABC} - g m_{AABB} \\ 
CA : 0 &= m_{A} - g m_{AABC} - g m_{ABAC} - m_{AAB} - g m_{AAAB} \\ 
CB : 0 &= -g m_{AABC} - g m_{ABAB} - m_{ABC} - g m_{AABB} \\ 
C^2 : 0 &= -m_{AAB} - 2 g m_{AAAB} - g m_{AABB} \\ 
A^3 : 0 &= m_{A}^2 - g m_{AAAAA} - 2 g m_{AAABC} + 2 m_{AA} - m_{AAAA} \\ 
A^2B : 0 &= m_{A}^2 - g m_{AAAAB} - g m_{AABBC} - g m_{AABCB} + m_{AB} - m_{AAAB} \\ 
A^2C : 0 &= m_{A}^2 - g m_{AAAAB} - g m_{AABBC} - g m_{AABCB} + m_{AB} - m_{AAAB} \\ 
ABA : 0 &= -g m_{AAAAB} - 2 g m_{ABABC} + 2 m_{AB} - m_{AAAB} \\ 
AB^2 : 0 &= -g m_{AAABB} - g m_{AAABC} - g m_{AABAC} + m_{AA} - m_{AABB} \\ 
ABC : 0 &= -m_{AABC} - g m_{AAABC} - g m_{AABCB} - g m_{ABABC} + m_{AB} \\ 
ACA : 0 &= -g m_{AAAAB} - 2 g m_{ABABC} + 2 m_{AB} - m_{AAAB} \\ 
ACB : 0 &= -m_{AABC} - g m_{AAABC} - g m_{AABCB} - g m_{ABABC} + m_{AB} \\ 
AC^2 : 0 &= -g m_{AAABB} - g m_{AAABC} - g m_{AABAC} + m_{AA} - m_{AABB} \\ 
BA^2 : 0 &= m_{A}^2 - g m_{AAAAB} - g m_{AABBC} - g m_{AABCB} + m_{AB} - m_{AAAB}  
\end{align*}
\subsection{The moments}
The following relations were found by solving some large set of SDE in Mathematica:
\begin{center}
\begin{align*}
    m_{AB} =& \frac{-m_1 - g m_2}{2 g}\\ 
m_{AAA} = &\frac{1 + 3 m_{1}^2 - m_2 + 9 g m_1 m_2}{
 3 g}  \\
m_{AAB}= &\frac{m_1 + g m_2}{6 g^2}\\ 
m_{ABC}= &\frac{2 - 3 m_{1}^2 - 2 m_2 - 9 g m_1 m_2}{6 g}\\ 
m_{AAAA}= &\frac{-7 g + 2 m_1 + 33 g^2 m_1 - 24 g m_{1}^2 + 18 g^2 m_{1}^3 + 9 g m_2 + 
   9 g^3 m_2 - 72 g^2 m_1 m_2 + 54 g^3 m_{1}^2 m_{2}}{27 g^3}\\
m_{AAAB}=& 
\frac{4 g - 5 m_1 + 3 g^2 m_1 - 3 g m_{1}^2 + 9 g^2 m_{1}^3 - 9 g m_2 - 
   9 g^3 m_2 - 9 g^2 m_1 m_2 + 27 g^3 m_{1}^2 m_2}{ 54 g^3}\\
m_{AABB}=& 
\frac{-8 g + m_1 - 6 g^2 m_1 + 6 g m_{1}^2 - 18 g^2 m_{1}^3 + 9 g m_{2} + 
   18 g^3 m_{2} + 18 g^2 m_{1} m_{2} - 54 g^3 m_{1}^2 m_{2}}{54 g^3}\\
m_{AABC}=&  \frac{-2 g - 2 m_{1} + 21 g^2 m_1 - 3 g m_{1}^2 - 18 g^2 m_{1}^3 - 
   9 g^3 m_{2} - 9 g^2 m_{1} m_{2} - 54 g^3 m_{1}^2 m_{2}}{54 g^3}\\ 
m_{ABAB}=& 
\frac{-8 g + m_1 - 15 g^2 m_1 + 24 g m_{1}^2 + 36 g^2 m_{1}^3 + 9 g m_{2} - 
   9 g^3 m_{2} + 72 g^2 m_{1} m_{2} + 108 g^3 m_{1}^2 m_{2}}{54 g^3} \\
m_{ABAC}=&
\frac{-2 g - 2 m_{1} + 30 g^2 m_1 + 6 g m_{1}^2 + 9 g^2 m_{1}^3 + 
   18 g^3 m_2 + 18 g^2 m_1 m_2 + 27 g^3 m_{1}^2 m_{2}}{54 g^3}\\
m_{AAAAA} =& 
 \frac{1}{ 81 g^4}\left (11 g - 4 m_{1} - 102 g^2 m_1 + 75 g m_{1}^2 + 144 g^3 m_{1}^2 - 
   90 g^2 m_{1}^3 - 15 g m_{2} + 90 g^3 m_{2} \right.\\
   &\left. + 225 g^2 m_{1} m_{2} + 
   135 g^4 m_{1} m_2 - 270 g^3 m_{1}^2 m_2 \right)\\
m_{AAAAB} = &
 \frac{1}{ 162 g^4}\left(-8 g + 7 m_1 + 3 g^2 m_1 - 3 g m_{1}^2 + 18 g^3 m_{1}^2 - 
   45 g^2 m_{1}^3 + 15 g m_2 \right. \\
   &\left.- 9 g^3 m_{2} - 9 g^2 m_1 m_2 - 54 g^4 m_1 m_2 - 
   135 g^3 m_{1}^2 m_2\right)\\ 
m_{AAABB} = &
 \frac{1}{162 g^4}\left(4 g + m_1 - 15 g^2 m_1 + 15 g m_{1}^2 - 36 g^3 m_{1}^2 + 
   63 g^2 m_{1}^3 - 3 g m_2\right. \\
   &\left.+ 45 g^3 m_2 + 45 g^2 m_1 m_2 + 108 g^4 m_1 m_2 + 
   189 g^3 m_{1}^2 m_2 \right)\\
m_{AAABC} = & 
 \frac{1}{162 g^4}\left(10 g - 2 m_1 + 3 g^2 m_1 - 3 g m_{1}^2 - 63 g^3 m_{1}^2 + 
   36 g^2 m_{1}^3 - 12 g m_2 \right.\\
   &\left.+ 45 g^3 m_2 - 9 g^2 m_1 m_2 - 135 g^4 m_1 m_2 + 
   108 g^3 m_{1}^2 m_2 \right)\\ 
m_{AABAB} = & 
 \frac{1}{162 g^4}\left(4 g + m_{1} - 15 g^2 m_{1} - 12 g m_{1}^2 - 36 g^3 m_{1}^2 - 
   18 g^2 m_{1}^3 - 3 g m_2 \right.\\
   &\left.- 9 g^3 m_2 - 36 g^2 m_1 m_2 - 54 g^4 m_1 m_2 - 
   54 g^3 m_{1}^2 m_2\right)\\
m_{AABAC} = &\frac{1}{162 g^4}\left(10 g - 2 m_1 + 30 g^2 m_1 - 30 g m_{1}^2 + 99 g^3 m_{1}^2 - 
   45 g^2 m_{1}^3 - 12 g m_2 \right.\\
   & \left.+ 18 g^3 m_2 - 90 g^2 m_1 m_2 + 27 g^4 m_1 m_2 - 
   135 g^3 m_1^2 m_2\right)
\end{align*}
\end{center}

\end{document}